\documentclass[acmlarge]{acmart}
\AtBeginDocument{%
  }

\usepackage{textcomp}
\usepackage{xcolor}
\usepackage{listings}   % for code formatting
\usepackage{xcolor}  
\usepackage{subfig}
\begin{document}

\title{STGen: A Novel Lightweight IoT Testbed for Generating Sensor Traffic for the Experimentation of IoT Protocol and its Application in Hybrid Network}

\author{Hasan Mahmood Aminul Islam}
\authornote{Authors contributed equally to this research.}
\email{hasan.mahmood@ewubd.edu}
\affiliation{%
  \institution{East West University}
  \city{Dhaka}
  \country{Bangladesh}
}
\author{Subrata Nath}
\authornotemark[1]
\affiliation{%
  \institution{East West University}
   \city{Dhaka}
  \country{Bangladesh}}
\email{shuvra.dev9@gmail.com}

\author{Mahamudur Rahman Maharaj}
\authornotemark[1]
\affiliation{%
  \institution{East West University}
   \city{Dhaka}
  \country{Bangladesh}}
\email{rahmanmehraj627@gmail.com}

\author{Nafis Shahriar}
\authornotemark[1]
\affiliation{%
  \institution{East West University}
   \city{Dhaka}
  \country{Bangladesh}}
\email{nafisshahriar003@gmail.com}

\author{Md. Khalid Mahbub Khan}
\authornotemark[1]
\email{khalid.khan@ewubd.com}
\affiliation{%
  \institution{East West University}
  \city{Dhaka}
  \country{Bangladesh}
}

\author{Riadul Islam}
\authornotemark[2]
\affiliation{%
  \institution{University of Maryland}
   \city{Baltimore Country, MD 21250}
  \country{USA}}
\email{riaduli@umbc.edu}

\renewcommand{\shortauthors}{Hasan MA Islam et al.}

\begin{abstract}

A Wireless Sensor Network (WSN) is a network that does not rely on a fixed infrastructure and consists of numerous sensors, such as temperature, humidity, GPS, and cameras, equipped with onboard processors that manage and monitor the environment in a specific area. As a result, building a real sensor network testbed for verifying, validating, or experimenting with a newly designed protocol presents considerable challenges in adapting a laboratory scenario due to the significant financial and logistical barriers, such as the need for specialized hardware and large-scale deployments. Additionally, WSN  suffers from severe constraints such as restricted power supply, short communication range, limited bandwidth availability, and restricted memory storage. Addressing these challenges, this work presents a flexible testbed solution named STGen that enables researchers to experiment with IoT protocols in a hybrid environment that emulates WSN implementations with the physical Internet through a dedicated physical server named \textit{STGen core}, which receives sensor traffic and processes it for further actions. The architecture of the STGen testbed adopts a modular and layered abstraction that facilitates experimentation, evaluation, and design decisions based on empirical studies. The STGen testbed is lightweight in memory usage and easy to deploy. Most importantly, STGen supports large-scale distributed systems, facilitates experimentation with IoT protocols, and enables integration with back-end services for big data analytics and statistical insights. In summary, STGen is an essential simulation tool for IoT researchers and developers, enabling them to accelerate their IoT experimentation through a command line interface (CLI) or a web interface. The key feature of STGen is the integration of real-world IoT protocols and their applications with WSN. Its modular and lightweight design makes STGen efficient and enables it to outperform other popular testbeds, such as Gotham and GothX, reducing memory usage by 89\%. While GothX takes approximately 26 minutes to establish a large topology with four VM nodes and 498 Docker nodes, STGen requires only 1.645 seconds to initialize the platform with 500 sensor nodes.
\end{abstract}

\keywords{Internet Protocol, IoT, Senor Traffic, Verification, Validation, ELK}

% \received{XXX}
% \received[revised]{XXX}
% \received[accepted]{XXX}

\maketitle

\section{Introduction}
Wireless sensor networks (WSNs) refer to networks of spatially dispersed and dedicated sensors that wirelessly collect and transmit data about physical or environmental conditions. These are similar to wireless ad hoc networks in the sense that they rely on wireless connectivity and the spontaneous formation of networks so that sensor data can be transported wirelessly. A typical sensor node consists of a radio transceiver, a microcontroller, sensor interfaces, and a power source, usually a battery or an embedded form of energy harvesting. These characteristics, along with limited computation capability of sensor nodes, restrict the node’s capabilities \cite{saginbekov2016testing}. However, these sensor nodes are equipped with various types of sensors, processing units, and wireless communication modules. It enables them to measure parameters such as temperature, humidity, pressure, motion, and more. The sensor nodes used in the WSN are connected to the sink node, which acts as a processing unit in the WSN system. In this work, we refer to the sink node as the STGen Core, which extracts and transmits sensor data for further processing. STGen uses a hybrid environment consisting of an emulated WSN and a physical STGen Core, with the emulated WSN effectively operating as a large network of sensor nodes. The STGen Core is a physical computational resource that facilitates communication between virtual sensor nodes and external systems. The STGen core facilitates the delivery of specific sensor data to client nodes over the public Internet upon request, enabling seamless interaction between the emulated network and real-world applications.

As Internet of Things (IoT) adoption continues to grow, the need for scalable and cost-effective sensor network simulation becomes increasingly critical for experimentation, verification, and validation of IoT systems. The concept of the IoT is wide-ranging from the commercial and technical point of view, such as smart cities, smart homes, pollution control, energy savings, smart transportation, and smart industries. However, setting up the IoT environment is time-consuming and expensive, especially with physical resources \cite{silmi2020wireless}. To address the issues, this article presents a lightweight testbed named STGen\footnote{https://github.com/rahmanmehraj182/Hybrid-IoT-Platform}, which generates enormous sensor traffic by mimicking an IoT-based WSN ecosystem. Therefore, STGen provides a promising approach to test research ideas using real-time traffic without interfering with the regular operation of the IoT ecosystem.

A crucial component in the development and prototyping process of network protocols is verification and validation. Evaluating a new feature involves balancing realism with cost considerations. Testbeds can offer valuable insight into how a new feature behaves, especially in WSN and IoT contexts. However, actual networking takes place on production environments, such as those found in campus, enterprise, and government environments. The research community uses simulators, emulators, and testbeds to conduct experiments in a controlled and reproducible environment. Without the need for physical devices and the complexity of real-world network deployment, the STGen provides seamless support for experimentation, verification, and validation. STGen could be a great tool for students and researchers to experiment with device integration and improve the performance of their IoT application services like real-time monitoring, traffic shaping, and edge-level data analytics. The most attractive and fascinating feature of the STGen tool is the ability to create and simulate thousands of connected IoT devices without having to configure and manage physical devices. In short, this work is intended for students, researchers, and developers who work on IoT protocols and applications and, therefore, need the verification and validation of their prototype through experiments on the IoT testbed, particularly in the early stage.

To the best of our knowledge, no prior research has introduced an emulated testbed tailored for hybrid network architectures, particularly those combining physical machines with large-scale virtualized sensor environments. The key contributions of this work are the following. \vspace{.1cm}
\begin{enumerate}
    \item Development of a lightweight, easily extensible, and customizable sensor traffic generator testbed for the experimentation of the IoT protocol.
    
    \item Easy-to-replicate and easy-to-deploy STGen platform in distributed networking systems.
    
    \item Platform independent, easy to control, and easy to configure through the STGen GUI and command line interface (CLI) in both Windows and Unix operating systems.

     \item Developed a hybrid simulation model of emulated Wireless Sensor Networks (WSNs), where virtual sensor nodes transmit data to a centralized STGen Core (sink node), which then streams the traffic to clients over the public internet.

    \item Integrating the ELK stack (Elasticsearch, Logstash, and Kibana) for real-time traffic ingestion and visualization, enabling dynamic and interactive graphing capabilities \cite{ngo2023new}.
\end{enumerate}

 Section \ref{background} presents the state-of-the-art of existing IoT testbeds and highlights how STGen improves scalability and flexibility. Section \ref{iot-testbed} outlines the core components of STGen and their role in supporting scalable simulations. Sections \ref{system_architecture} and \ref{key-feature-system-architecture} present the system architecture and technical details about the serialization of data, the transmission in real time, and the design of interoperable middleware with the STGen. The setup of the environment in \ref{environment} describes the tools needed to deploy STGen. The Results section \ref{observations} presents performance metrics that demonstrate the efficiency of the testbed. The comparison Section \ref{discussion} briefly describes the effectiveness of STGen in simulating large-scale IoT environments compared to recent testbeds. The Related Work compares STGen with other IoT testbeds. Finally, Section \ref{conclusion} concludes the work with the future direction of this work.

 \begin{table*}[!t]
\caption{Comparison of IoT Testbed Platforms}
\label{table_example}
\centering
\resizebox{\textwidth}{!}{%
\begin{tabular}{cccccc}
    \toprule
   Features  & Gotham  &GothX & IoT Flock  & PatrIoT & STGen \\
    \midrule
        Lightweight & No & No & Yes & No & Yes \\
        Memory Footprint & Very High & Very High & High & n/a & Low \\
        Sensor Node Initialization Time & Very High & Very High & n/a & High & Low \\
        Deployment test on Single Machine & Up to 140 nodes & Up to 450 nodes & n/a & n/a & Up to 6k nodes \\
        Real Time Log Analysis & No & No & No & No & Yes \\
        CLI support & No & No & No & Yes & Yes \\
        Stress Testing Capability & Lacks support & Lacks support & Yes & Hardcoded & Yes \\
        Message Rate Customization & No & Yes & Yes & Hardcoded & Yes \\
        Protocol Customization & No & No & No & No & Open for extension \\
        Modularized & n/a & n/a & No & Yes & Yes \\
        Simplicity & Not user-friendly & Complex & Complex & Hardcoded & Simple and intuitive \\
        Node Count Customization & Hardcoded & Yes & Yes & Yes & Yes \\
        Data Archival & Yes & Yes & Yes & Yes & Yes \\
        Reproducible Sensor Configuration & Yes & Yes & Yes & Yes & Yes \\
        Application Level Logs & Yes & Yes & Yes & Yes & Yes \\
        GUI & Yes & Yes & Yes & No & Yes \\
        OpenSource Project & Yes & Yes & Yes & Yes & Planned release \\
    \bottomrule
\end{tabular}
}
\end{table*}

\section{Motivation and Scope}
\label{motivation}
The key motivation and vision of this work is introducing a novel and lightweight IoT testbed for the IoT research community, in particular those who are working on the development of a novel IoT protocol and require the verification of their prototype at the early stage of their development, as well as the processing of the voluminous sensor traffic. Building an interoperable and universal testbed for the IoT with different types of physical devices/sensors embedded in the microcontrollers is very expensive. The research community working on the development of IoT protocols requires a testbed for the verification and validation \cite{minani2024systematic} of the prototype, particularly in the early stages of development. Early-stage researchers need significant time to become familiar with the microcontroller, to learn how to embed different types of sensors in the microcontroller, and to experiment with the whole ecosystem. However, researchers and students need a reasonably affordable and efficient environment to evaluate the ongoing development of IoT protocols. In response to this constraint, this research introduces STGen, a lightweight, scalable, and easy-to-use tool designed to address the challenges effectively. Due to its lightweight architecture, STGen can emulate thousands of virtual sensor nodes efficiently, making large-scale testing feasible on commodity hardware. Its design inherently adapts to the available system resources, supporting a wide range of IoT scenarios without the burden of managing physical infrastructure. Furthermore, the proposed architecture of the STGen not only simplifies the simulation but also supports scalable and customized environments for experimenting with various IoT protocols and opens a great opportunity for back-end services, such as big data analytics.

\section{State of the Art}
\label{background}

In recent years, several experimental testbeds have been developed to support a large-scale IoT environment for the experimentation of IoT protocols and their applications. For example, SmartSantander \cite{sanchez2014smartsantander}, Phynetlab\cite{falkenberg2017phynetlab}, and FIT IoT-LAB \cite{adjih2015fit} have advanced the field by providing researchers with extensive sensor networks. However, these testbeds often come with limitations in node accessibility and complex configurations. For instance, the main limitations of SmartSantander are its battery life and limited user reprogramming options, while FIT IoT-LAB focuses on providing access to bare metal but at the expense of scalability. In contrast, STGen is a highly flexible and scalable platform that enables the configuration of thousands of nodes with ease. This approach not only minimizes the cost of developing a physical IoT platform but also maximizes the efficiency of data transport across thousands of nodes, making it a state-of-the-art solution for developing an IoT protocol.

\begin{figure}[!t]
  \centering
  \includegraphics[width=.95\textwidth]{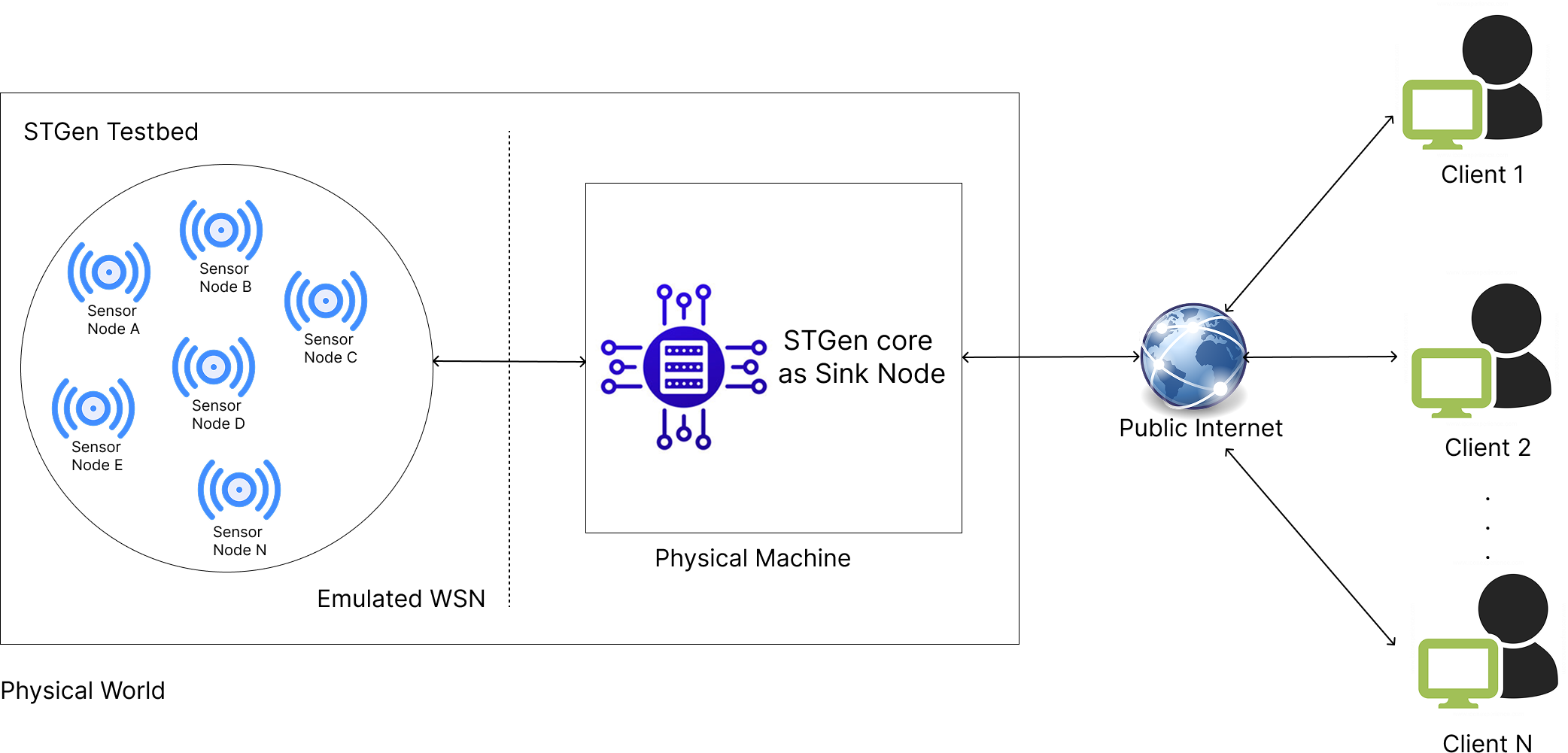}
  \caption{Hybrid model of STGen testbed integrating an emulated WSN with a sink node, enabling remote clients to receive IoT traffic via the public internet.}
  \label{stgeco}
\end{figure}

The authors in \cite{tsakalidis2023design} introduce an open-source IoT testbed constructed on the OpenHAB platform, incorporating diverse components, including wireless interfaces, Z-Wave, ZigBee, Wi-Fi, 4G-LTE, and infrared, as well as numerous sensors. This testbed concentrates in delivering real-time monitoring and sustaining a robust, persistent layer. IoT-flock \cite{ghazanfar2020iot} is also an open-source framework for the generation of IoT traffic to develop security solutions for smart homes based on real-world IoT devices. IoT-flock supports the widely used IoT application layer protocols, MQTT and CoAP. However, it does not include memory‑footprint measurements or performance benchmarks across varying node counts, nor does it discuss its limitations or compare resource efficiency with other testbeds. Similar work has been done in this paper \cite{shahid2020generative} but with a different strategy to generate different data categories. Another work similar to STGen, the Patriot \cite{bures2021patriot}, has been developed to provide a flexible IoT system testbed that provides scalability from a physical testbed to an emulated environment to an experimental IoT testbed. The dependency on  JUnit 5 of PATRIOT requires the hardcoding of simulation settings in individual test classes. Conversely, STGen provides a command-line interface enabling users to dynamically adjust simulation settings during execution, hence enhancing customisation and permitting seamless incorporation into automated workflows.

According to the author's study \cite{papadopoulos2013adding}, testbeds and simulators serve as complementary validation tools for successful real-world deployment experiments. The STGen platform functions as an IoT testbed capable of simulating wireless sensor networks, with users controlling network traffic patterns through configurable CLI parameters. A clean and modular testbed with a proper coding structure improves maintainability, simplifies modification, and guarantees flexible usability for researchers. This underscores the importance of a virtual testbed that provides reusability, as ``reinventing the wheel'' in such circumstances is inefficient and unnecessary.

The authors of this research \cite{buratti2015testing} have tested various communication protocols, including ZigBee, IPv6, and SDWN (which employs a centralised network layer protocol), using the EuWIn Platform. ZigBee and IPv6 are two of the most common standards for IoT applications; more recently, another approach based on the software-defined network (SDN) paradigm has been proposed. However, the authors experimented on very few nodes, which showed the limitations of the physical testbeds.

The authors in \cite{pan2015internet} have developed a physical IoT testbed by integrating sustainable and energy-efficient features and analyzing energy consumption trends related to environmental conditions and occupancy rates. The lessons learnt from this IoT testbed indicate that estimating environmental and human-related parameters is crucial to enhancing sustainability. However, the work needs more details on the scalability of large-scale IoT systems; it also requires information on how to simulate the testbed or process data, limiting its applicability in broader research scenarios and reproducibility.

Gothx \cite{poisson2024gothx}, a flexible traffic generator, and Gotham \cite{saez2023gotham}, a reproducible IoT testbed, have been developed to create IoT data sets with legitimate and malicious traffic. They support validation across three different communication protocols (MQTT, Kafka, and SINETStream). The STGen platform is complementary to Gothx but is more efficient in terms of memory usage and runtime performance. The MakeSense \cite{jiang2020makesense} is a flexible and easy-to-install testbed for real-life large-scale IoT applications that was introduced by the researchers in \cite{jiang2020makesense}. In this case, real-life testing was performed on the designed testbed only for two IoT applications: homes and offices. However, other promising IoT applications are not evaluated here, whereas the modules of this testbed are cost-minimizing.

This work \cite{alsukayti2020multidimensional} presents an analogy that shows how modular design could make a testbed efficient, scalable, and extensible. Various Internet of Things (IoT) properties were taken into account during the development of this framework. Although this study covers the architecture and use case, it does not go into detail on the testbed system, such as its performance or memory footprint. However, the primary goal of this article was to provide the groundwork for developing a lightweight testbed system. Due to a lack of comparison with other testbeds, this is merely a simulation with a small number of sensor nodes.

Current IoT research trends show that researchers are increasingly striving to improve existing IoT protocols to enable device cognition and intelligence, which is why researchers need to rely on experimentation-based prototyping for proof-of-concept development. According to \cite{papadopoulos2017thorough}, the majority of concepts that were validated by simulation were later confirmed using experimental testbeds. The primary benefit of testbeds is the reproducibility of several instances of the same situation by researchers who did not design the testbed themselves \cite{chernyshev2017internet}. 

OpenTestBed \cite{munoz2019opentestbed} is an open-source IoT testbed to evaluate and benchmark IoT solutions. This testbed needs standard components such as Raspberry Pi single-board computers, OpenMote B low-power wireless devices, and glass domes. Nevertheless, STGen is complementary to OpenTestBed but without ``heavy" installations with dedicated wiring, switches, servers, etc. However, a comprehensive summary of STGen with the most recent existing simulators, emulators, and testbeds for different IoT application environments is shown in Table \ref{table_example}.

\begin{figure}[t]
  \centering
  \includegraphics[width=1\textwidth]{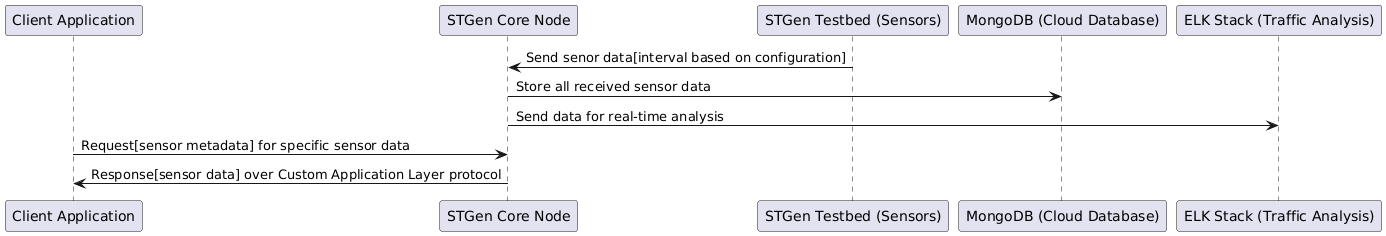}
  \caption{Sequence diagram of STGen testbed illustrating client applications requesting the STGen Core Node with sensor metadata and receiving sensor data over a custom application layer protocol. The STGen core node collects sensor data from the testbed at configurable intervals, stores it in a cloud database, and forwards it to the ELK Stack for real-time traffic analysis.}
  \label{stgenseq}
\end{figure}

\section{STGen as Hybrid Network Model}
\label{iot-testbed}

In STGen, the emulated Wireless Sensor Network (WSN) nodes are not constrained to a single physical machine. These software-defined nodes can be distributed across multiple systems and communicate with the STGen Core either via wired channels (such as Ethernet or LAN switches) or wirelessly (e.g., Wi-Fi). This architectural flexibility enables the formation of spatially separated WSN clusters that still maintain synchronized communication with the core. The STGen Core itself is decoupled from the emulated node's host systems, thus enabling a hybrid network model—one that integrates both wired and wireless communication modalities. Building on this foundation, STGen serves as an IoT testbed, as illustrated in Figure \ref{stgeco}, offering a powerful platform for traffic generation and protocol evaluation in ybrid IoT networks. The testbed is designed to simulate thousands of sensors and actuators, which, in the real world, are typically resource-constrained \cite{shelby2022rfc}, that is, are powered/operated by batteries and have a limited amount of storage and processing capabilities. Therefore, communications in IoT are sensitive to time. In such scenarios, the User Datagram Protocol (UDP)~\cite{eggert2017rfc} is the appropriate transport protocol for the communication of sensor data in WSN. Within Wireless Sensor Networks (WSNs), sensor nodes are generally deployed within the boundaries of a private network. As a result, direct communication with external systems is generally mediated through a designated sink node. In the proposed hybrid network model, a physical machine is designated as the STGen core, which assumes the role of the sink node, effectively offloading the computational and communication burdens from the emulated sensor nodes.

The fundamental idea of STGen aligns with the well-known phrase, ``You cannot see the forest for the trees". Despite its complexity, STGen aims to provide a platform for researchers or students engaged in the development of IoT protocols that require prototype verification during the initial or intermediate stages of development. The key components of the STGen testbed are the following:

%%%%% the proverb is repetitive. So removed one

\begin{itemize}
    \item \texttt{Sensors:} These are emulated sensor devices that generate sensor traffic.
    \item \texttt{STGen core as Middleware/Sink Node:} This component collects sensor traffic and stores data for future inspection.  
    \item \texttt{Transport Protocol:} This design choice is crucial in constrained environments where resources are scarce. Unlike TCP \cite{eddy2022rfc}, which ensures reliable data delivery using a connection-oriented approach, UDP is a connectionless protocol and requires less overhead compared to TCP. Since most of the IoT protocols are developed on top of UDP, STGen employed a custom, extensible application-layer protocol over UDP to facilitate flexible and lightweight communication.
    
    \item \texttt{Real-Time Log Analysis:} A key component of the STGen testbed is its ability to support real-time traffic visualization. Researchers can experiment with their protocols using the testbed while leveraging this feature to monitor and measure performance dynamically through visualization graphs with the ELK stack. Even though big data analytics frameworks such as Apache Hadoop and Apache Spark are not integrated with STGen, its modular design allows the possibility of future integration. This makes it possible for researchers to enhance the capabilities of the testbed for processing massive amounts of data, which, in turn, enables more in-depth analysis and optimization of Internet of Things protocols.
\end{itemize}

STGen enables independent simulation of its STGen core server, sensors, and client applications via both the Web UI and CLI. As illustrated in Figure \ref{stgenLauncher}, the STGen Core can be initialized by opening a designated port, allowing the client application to communicate with it, along with defining the sensor data receiving port for the incoming sensor traffic from the sensor nodes. Since sensor nodes operate independently, users must specify the STGen Core address to ensure seamless communication between the locally running sensors and the STGen Core. Additionally, the simulation allows sensors to be configured for a pre-defined duration, allowing control over the number of nodes and their respective transmission behaviors. Similarly, as shown in Figure \ref{stgenLauncher}, the client application can be executed with a defined STGen Core address to establish communication. Users must also specify an archive directory to store sensor data, facilitating data retrieval based on specific sensor identifiers of interest. This modular approach improves the flexibility and scalability of IoT simulations within the STGen framework. As shown in Figure \ref{stgenseq}, the sequence diagram details the step-by-step communication among the key components of the STGen architecture.

\section{Key Requirements of IoT Simulation}
The STGen testbed provides a scalable and efficient hybrid environment for large-scale IoT research and experimentation. It incorporates the following key capabilities \cite{almutairi2024advancements}:

\begin{description}

    \item[Scalability Testing:] STGen emulates extensive IoT networks through a simulated Wireless Sensor Network (WSN), facilitating scalable and adjustable sensor deployments for the research, testing, and validation of IoT communication protocols and data processing.
    
    \item[Cost-Effective Deployment:] By providing a virtualized environment, STGen eliminates the need for extensive physical hardware. Researchers can configure and test IoT protocols, sensor interactions, and data transmission efficiently without additional infrastructure investments.

    \item[Simulation Tools and Platforms:] STGen offers platform-agnostic support, facilitating smooth functionality across various environments. The testbed's cross-platform interoperability allows deployment and operation on several operating systems, enhancing flexibility and usability for researchers and developers.
    
    \item[Latency and Throughput Analysis:] STGen allows researchers to simulate network failures, apply controlled delays, and examine throughput fluctuations to assess system resilience and fault tolerance. Metrics including latency, packet loss, and performance trends are monitored and visualised through the ELK stack for real-time analysis.
    
    \item[Lightweight Communication and Protocol Flexibility:] STGen employs a custom application layer protocol on top of UDP for efficient and low-latency communication. It utilizes BSON encoding for packet payloads, guaranteeing optimized data transport and interoperability across various IoT applications.
    
    \item[Prototyping and Validation:] The testbed enables rapid prototyping of IoT systems by supporting virtual sensor deployments and application-layer protocol testing. The STGen facilitates the validation of sensor data transmission, processing workflows, and system integrations before real-world deployment.
    
    \item[Simulation of a real-world scenario:] STGen provides configurable environmental conditions. Researchers can utilize the traffic tool to simulate packet loss, generate latency, set bandwidth limits, and more to test an IoT system in different environments.
    
    \item[Big Data Analytics and Storage:] The STGen core efficiently aggregates and stores large quantities of sensor data in MongoDB for further analysis, while logs are maintained on the local disk for debugging and performance evaluation. These large-scale datasets can be fed into machine learning models for big data analytics.

    \item[Training and Educational Use:] STGen provides an intuitive interface for configuring and visualizing simulations, allowing students and developers to experiment with IoT systems and understand network behaviors without requiring physical hardware.
    
\end{description}

\section{STGen System Architecture}
\label{system_architecture}

The architecture of the STGen testbed is designed with a strong focus on scalability and customization, ensuring that each module can be individually deployed and operated independently without dependencies on other components. Each layer is built to ensure that the modules remain decoupled, preventing dependencies that could impact each other. Each module has a well-defined and specific responsibility, allowing for seamless integration, flexibility, and efficient evaluation of IoT prototypes. The key features of the STGen system architecture are discussed below.

\subsection{Core Components}
\label{core-components}

The core components of STGen considered at the early stage of prototyping are described as follows:

\begin{description}
    \item[Sensors:] The sensor nodes of the STGen platform resemble real-world IoT sensors that generate values of a wide variety of variables, such as temperature, GPS, and camera data, at varying rates. One of the key advantages of STGen is its modular and extensible architecture, which allows seamless integration of new sensor types with minimal effort. The framework is designed to accommodate diverse IoT applications by providing a \texttt{plug-and-play} mechanism for sensor modules. Adding a new sensor type requires only two fundamental steps: 
    \begin{description}
        \item[Defining the Sensor’s Default Behavior:] Implement the logic for generating data and defining the transmission characteristics of the sensor.

        \item[Configuring its transmission parameters:] Use the current configuration system to dynamically manage data transmission rates.
    \end{description}
     
    \item[STGen Core:]
        The STGen core is the central component of the STGen design, located between the client and the sensor nodes. It includes a node registry that records the states of both the client and the sensor nodes. STGen core acts as a sink node, accumulating data from the sensor nodes and facilitating delivery to the client application through the public Internet. In addition, the STGen core stores the collected sensor data and application logs on the local hard disk and periodically stores the sensor data in MongoDB at runtime for further use.   
    
    \item[Client Application:] Client applications of the STGen platform are connected to the STGen core to interact indirectly with the sensor nodes. Although sensor nodes do not recognize client nodes, their main task is to transmit data to the STGen core. Clients can retrieve the traffic from the STGen core by submitting a request for sensor nodes. The objective of this separation of concerns is to ensure that the core handles the complexities of sensor data management, allowing clients to focus just on the sensor data itself. This architecture accommodates numerous clients who are interfacing with the identical STGen core.

    \item[BSON Serialization for Compact Data:] In contrast to traditional testbeds, the primary advantage of STGen lies in its simplicity, with each configuration meticulously designed to optimize access patterns. To enhance efficiency in the STGen testbed, BSON (Binary JSON) serialization has been employed, prioritizing payload compactness and minimizing bandwidth consumption. BSON, as a binary encoded representation of JSON, offers a significantly reduced payload size – our experimental benchmarks demonstrate that BSON achieves approximately 76\% of the size of an equivalent JSON payload. This reduction in payload size directly contributes to faster data traversal across the network, a critical factor for real-time IoT communication systems. Although BSON introduces higher serialization and deserialization overhead compared to JSON, its compact format effectively mitigates network latency and bandwidth usage. This ensures efficient data transmission, even in resource-constrained IoT environments, making BSON a suitable choice to optimize performance in such scenarios.
    \end{description}

\subsection{Flexible Configuration}
\label{system-architect}

The STGen testbed can operate in the command-line interface (CLI) mode and WEB server mode, providing different configuration options. The Web-based GUI provides a convenient interface for configuring and accessing the testbed smoothly without prior knowledge of the CLI mode. The Web interface converts user choices into API calls, and the Web server transforms those API calls into CLI commands to execute the desired setup and control actions on the testbed. However, STGen can be described as a user-friendly tool that features very simple and intuitive parameter options. STGen can be a very useful tool for researchers and developers in modeling complicated settings easily because of the adaptability of the STGen tool in a broad range of IoT research domains.

\begin{figure}[htbp]
  \centering \includegraphics[width=0.8\textwidth]{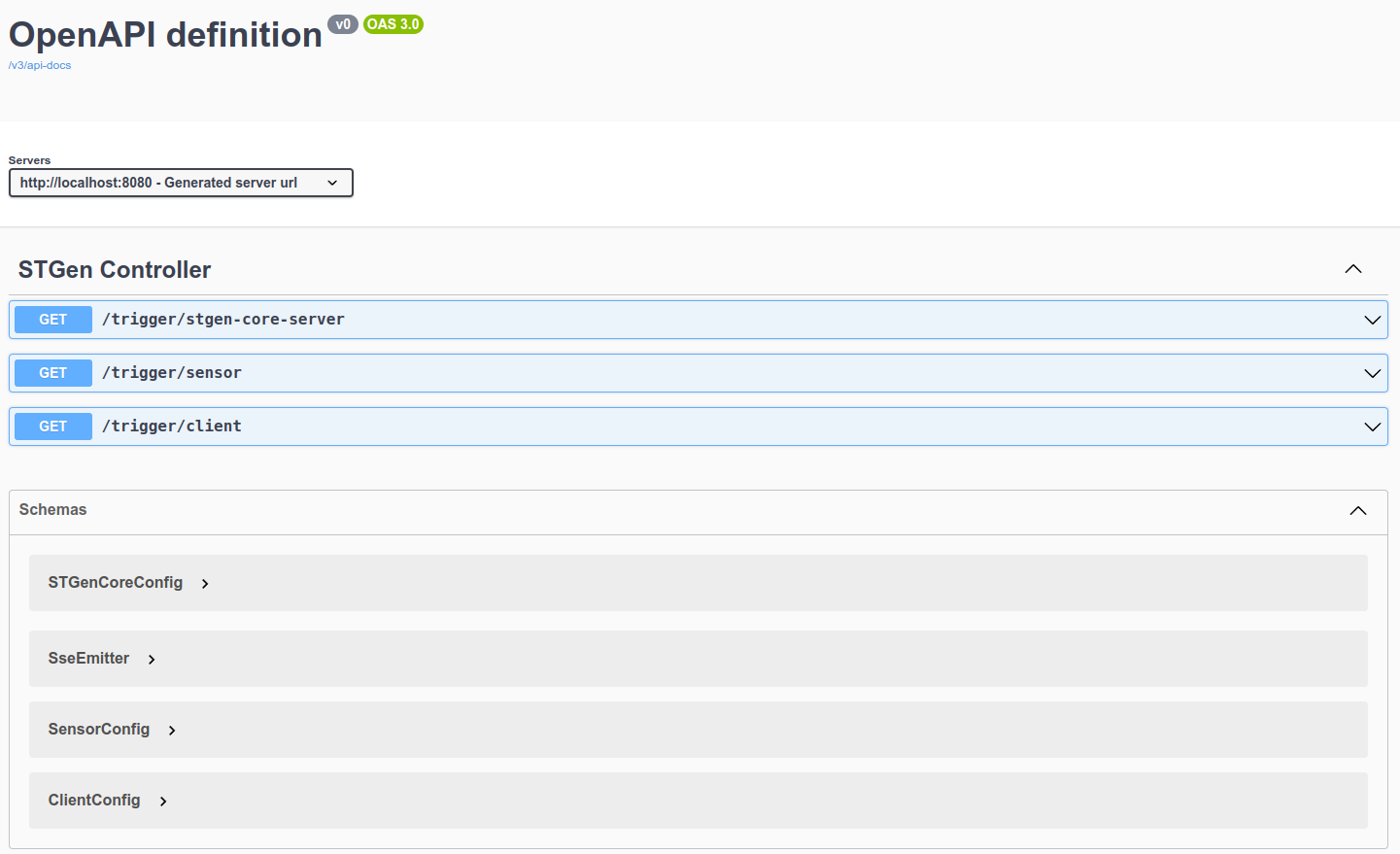}
  \caption{The STGen OpenAPI definition displayed in the Swagger UI (/swagger-ui/index.html).}
  \label{openAPI}
\end{figure}

To further enhance usability and automation, users can run the backend service that exposes these REST endpoints, converting REST API query parameters into the appropriate CLI commands—an approach particularly beneficial for those who prefer to avoid direct terminal command execution. This backend service simplifies and automates processes such as sensor, client, and core creation, reducing manual intervention and enhancing reproducibility. Additionally, API calls stream real-time application logs directly to the browser when utilizing the front-end service, providing transparent insight into ongoing operations. For researchers relying solely on the REST API endpoints, we have defined a comprehensive OpenAPI specification \ref{openAPI}, which is available through the Swagger UI at /swagger-ui/index.html. This includes detailed documentation, interactive testing capabilities, and version control features, further facilitating integration, reproducibility, and ease of use. Moreover, for those who prefer direct interaction with the CLI, all available CLI commands are provided in the appendix section for reference.

\subsection{Real-Time Data Transmission and Customization}

The primary issue of IoT networks lies in the varied characteristics of sensors, actuators, and devices. This encompasses ``heterogeneity of devices," ``heterogeneity in data formats," and ``interoperability issues arising from heterogeneity" \cite{noaman2022challenges, sowe2014managing, moon2019heterogeneous}. When forecasting heterogeneous IoT data values, temporal and geographic dependencies become particularly evident and are influenced by numerous additional factors, as they are dependent on specific situations. Our system incorporates a default transmission rate that can be modified through input parameters, allowing researchers to customize the rate for particular sensor types or applications, hence improving the versatility of the testbed.

% Although several IoT testbeds are available, many standard frameworks presume a uniform data transfer rate, failing to represent real-world dynamics.

\section{Key features of STGen Architecture}
\label{key-feature-system-architecture}
The key features offered by the STGen architecture are summarized as follows:

\begin{description}
    \item[STGen Core as the Core Processing Unit:] The framework simulates the behavior of sensor nodes publishing data to the STGen core, which acts as a rendezvous of the simulation ecosystem. The core serves as the portal between the client application and the sensor nodes. The client application carries out the communication with the STGen core, while the core stores and processes the sensor data.

    \item[Dual-Archiving for Redundancy and Scalability:] As the STGen core is ephemeral, it is essential to aggregate the data transmitted by the different sensor nodes in a storage system. For this, the STGen architecture performs two distinct tasks: first, distributing the data to MongoDB for efficient retrieval and analysis, and second, ensuring that the sensor data is aggregated locally. This dual-archiving strategy ensures redundancy and facilitates access in real-time or temporal analysis later on. The network traffic log is indexed and stored in Elasticsearch for real-time visualization and meaningful insights.
  
    \item[Decoupling Sensors and Clients through STGen Core:] The scalability of the simulation environment is achieved through the decoupling of different components of the STGen Core, which is briefly summarized as follows:

    \begin{description}
        \item[Decoupling for Focused Sensor Operation:]
        By separating the client applications from the sensor nodes, the STGen design allows the sensors to concentrate on data collection while the middleware (an instance of the STGen core unit node) processes, routes, and aggregates the data locally or in a cloud database.
        % can be replaced by cloud. as we used MONGODB,
    
        \item[Middleware as the Communication Gateway:] The STGen core instance acts as middleware between the sensor nodes and the client applications. The core manages all communication as the sensor nodes are resource-constrained. The client applications interact only with the sensors through the STGen core. 
        
        \item[Scalability through Decoupling:] By offloading the sensor-generated data to the different storage systems by the STGen core, the system achieves high scalability and can simulate millions of sensor nodes from the distributed testbed without increasing the complexity of client applications. This approach supports the deployment of large-scale IoT networks, reducing the operational load on a particular cluster.
    \end{description}
\end{description}

\begin{figure}[t]
  \centering
  \includegraphics[width=.95\textwidth]{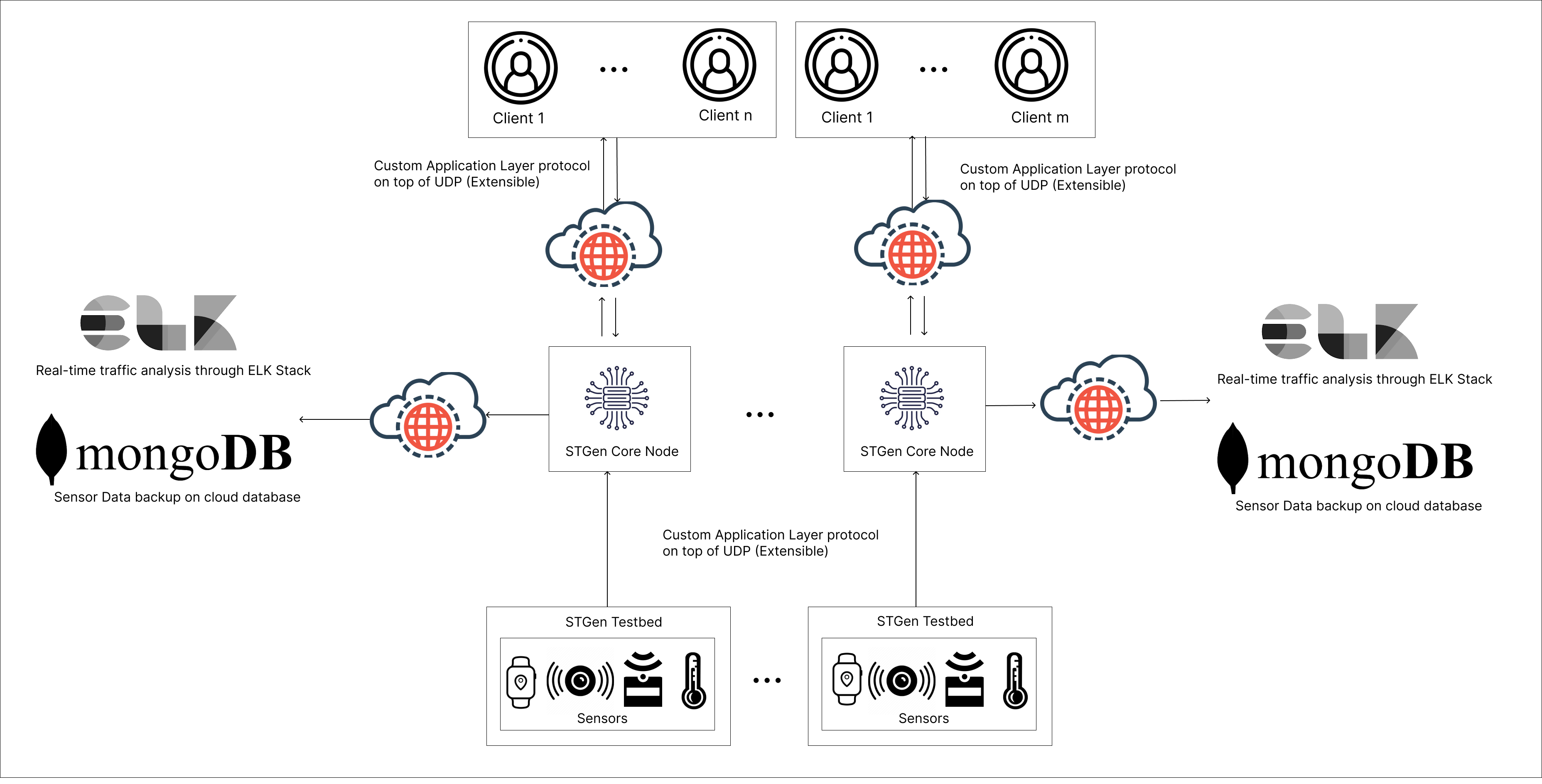}
  \caption{Distributed STGen testbed setup for experimenting with various IoT protocols and big data analytics on sensor data. Emulated testbeds feed sensor data to the STGen Core, which processes and provides real-time traffic analysis via the ELK Stack. Client applications access data using a custom UDP protocol, with sensor data backed up to a MongoDB cloud database.}
  \label{stgen-eco-overview}
\end{figure}

\section{Distributed Setup of STGen Testbed}
\label{environment}

\begin{table*}[htbp]
\centering
\resizebox{\textwidth}{!}{%
    \begin{tabular}{cc}
     \toprule
        \textbf{Parameter} & \textbf{Description} \\
     \midrule
        STGen Core IP &  IP address of the server to connect the sensors. \\
        STGen Core Sensor Incoming Port & Port number on the STGen Core to which sensors send data. \\
        Simulation Time & Total simulation time in seconds. \\
        Sensor Config & Configuration for each sensor type in the format \texttt{sensor\_type:number\_of\_sensors:data\_rate}. \\
        Sensor Type  & Type of sensor available in the simulation (\texttt{switch, temp, gps, camera}) \\
        Number of Sensors & Number of sensors of this type to simulate. \\
        Data Rate & Percentage of the base transmission rate at which the sensor generates and transmits data\\
     \bottomrule
    \end{tabular}
    }
    \caption{Simulation Parameters for STGen Sensor Testbed}
    \label{tab:sim_params}
\end{table*}

\begin{table*}[htbp]
    \centering
    \begin{tabular}{cc}
     \toprule
        \textbf{Parameter} & \textbf{Description} \\
     \midrule
        STGen Core IP &  IP address where the server listens for incoming connections. \\
        Sensor Port &  UDP port where the server receives sensor data from simulated sensors. \\
        Client Port &  UDP port where client applications request to receive sensor data. \\
        Simulation Time & Total simulation time in seconds. \\
    \bottomrule
    \end{tabular}
    \caption{Simulation parameters for STGen Core}
    \label{tab:stgen_server_params}
\end{table*}

\begin{table*}[htbp]
    \centering
    \begin{tabular}{cc}
     \toprule
        \textbf{Parameter} & \textbf{Description} \\
     \midrule
        Log Directory & Specifies the directory for storing output sensor logs on the client side. \\
        STGen Core IP & The IP address of the STGen Core. \\
        STGen Core Client Port & The port number of the STGen Core where the client sends requests. \\
        Simulation Time & The total duration of the simulation in seconds. \\
     \bottomrule
    \end{tabular}
    \caption{Simulation Parameters for STGen Client}
    \label{tab:stgen_client_params}
\end{table*}

This section describes the recommended environments for the deployment of sensor nodes for simulation and how to configure those nodes in the simulation. In addition, this section presents how to benchmark the STGen platform in both the local machine and the distributed networking system.

Tables \ref{tab:sim_params}, \ref{tab:stgen_server_params}, and \ref{tab:stgen_client_params} delineate the respective parameters for the STGen Sensor Testbed, the STGen Core, and the STGen Client.

\begin{description}
    \item[Deployment of Sensor Nodes:] The STGen environment currently supports five different types of sensors for deployment, such as temperature, humidity, camera, GPS, and switch (on / off). These sensors are selected by CLI or the Web interface. The number of sensors for the experiment depends on the configuration of the workstation system.
   
   \item[Hardware Configuration of the Workstation:] The system configuration of the workstation used in the experiment is the following:
   
   \begin{enumerate}
        \item \texttt{Processor:} AMD Ryzen 3600X
        \item \texttt{RAM:} 36 GB
        \item \texttt{SWAP Memory:} 6 GB
        \item \texttt{GPU:} NVIDIA 4060 Ti
        \item \texttt{Operating System:} Ubuntu 24.10
    \end{enumerate}

    \item[Traffic Control Configuration:]
       The \texttt{tc} (Traffic Control) tool \cite{ubuntu_tc} was used to simulate network conditions that mimic real-world environments. The following configurations were applied:

     \begin{itemize}
        \item \texttt{Bandwidth Limitation:} Unbounded / 100 Kbps / 10 Kbps
        \item \texttt{Packet Loss Simulation:} 0\%, 1\%, 5\%, 10\% packet loss using \texttt{netem}
        \item \texttt{Latency:} 10 ms maximum latency for packets
      \end{itemize}

\lstset{
    basicstyle=\ttfamily\small,
    frame=single,
    breaklines=true,
    breakatwhitespace=false, 
    postbreak=\mbox{\textcolor{red}{$\hookrightarrow$}\space}
}

\begin{lstlisting}
$ sudo tc qdisc add dev lo root netem rate 10kbps loss 10%
\end{lstlisting}

The commands stated above illustrate how bandwidth limitation and packet loss were applied in the experiment.

\begin{figure}[htbp]
  \centering
  \includegraphics[width=0.7\textwidth]{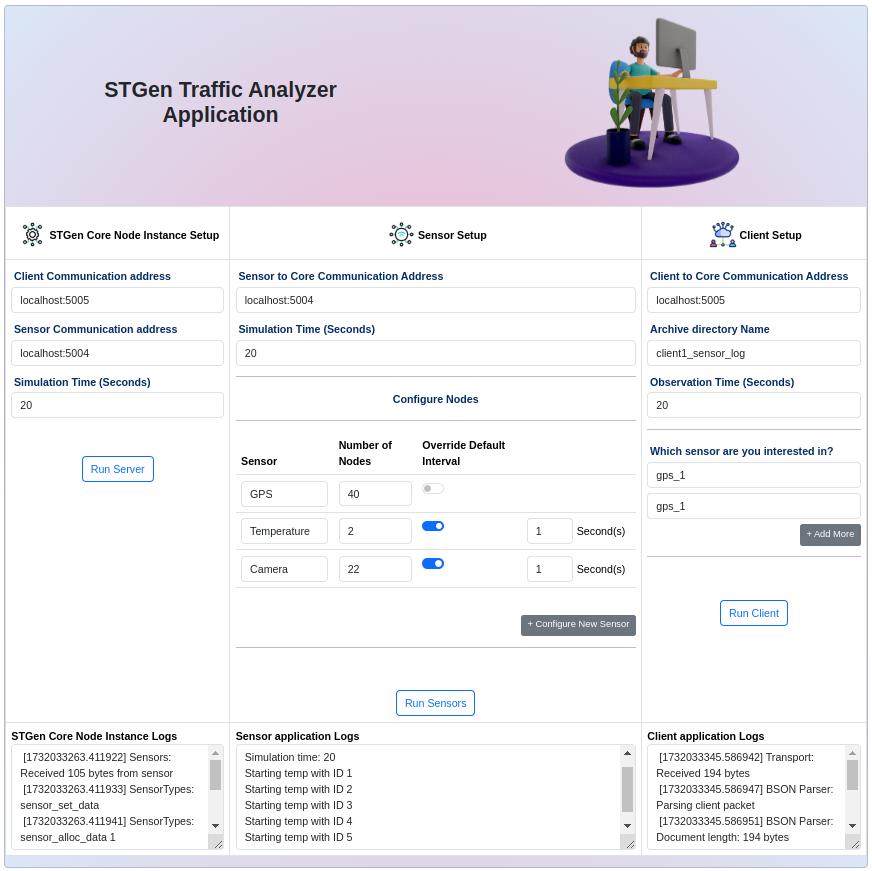}
  \caption{STGen Testbed Launcher Web UI.}
  \label{stgenLauncher}
\end{figure}

\begin{figure*}[htbp]
  \centering
  \includegraphics[width=\textwidth]{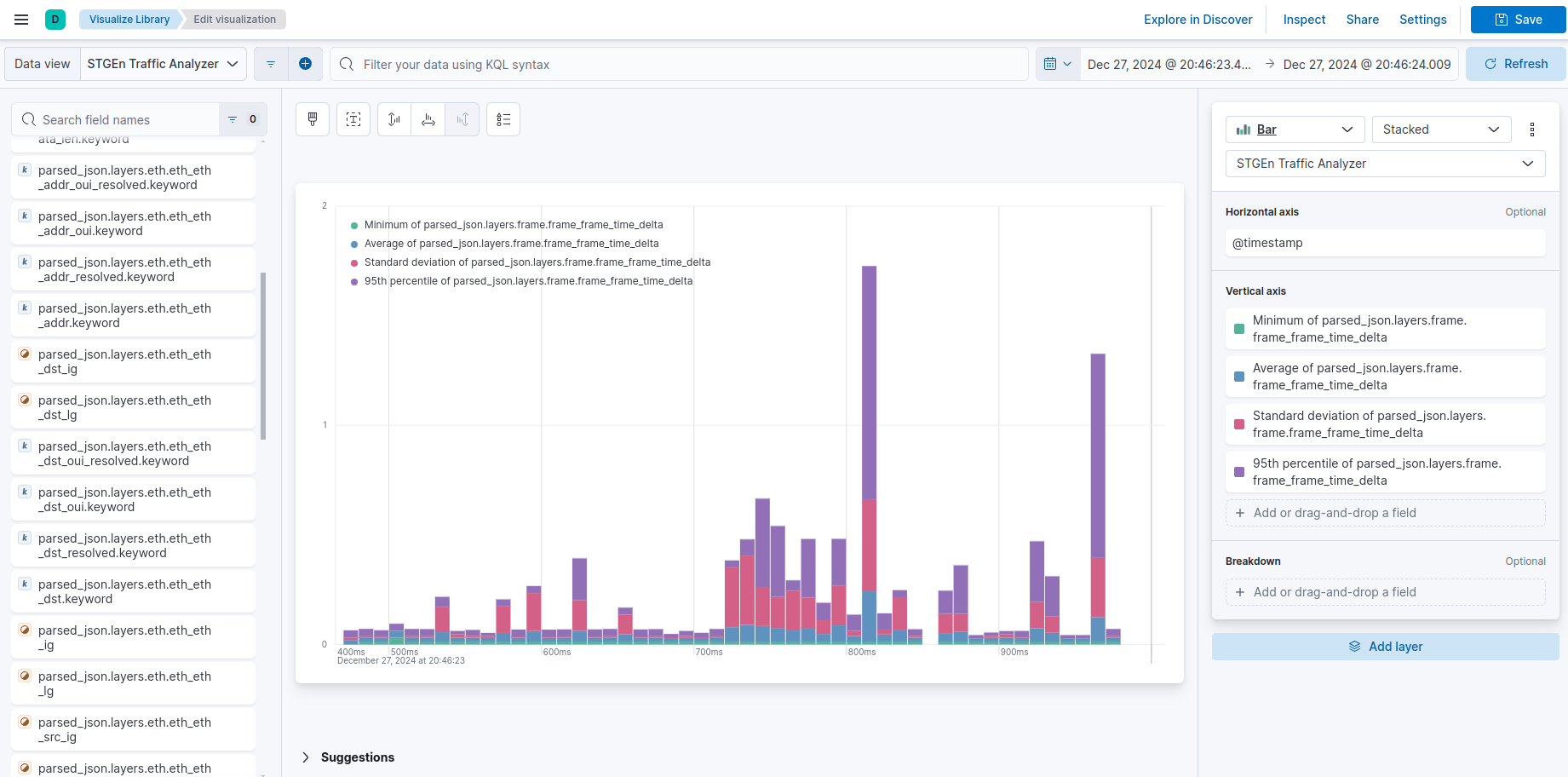}
  \caption{Real-Time Delay Statistics of Network Traffic in STGen Traffic Analyzer – This graph shows the minimum, average, standard deviation, and 95th percentile of frame time delta (delay) over time, visualized using the ELK stack for real-time traffic analysis.}
  \label{elk}
\end{figure*}

\item[Testbed Simulation with Web UI:]
A web-based user interface (WebUI), as shown in Figure \ref{stgenLauncher}, has been implemented to set up and control the experiments with the STGen testbed. The WebUI makes the testbed more user-friendly compared to the CLI mode for monitoring sensor nodes in the IoT network. The testbed also supports the control of the system. Users can assign values to different parameters and start simulations using the user interface and trigger the direct streaming of the process logs to the interface, which improves transparency and makes debugging easier by providing real-time feedback.

\item[Realtime Network Log Analysis:]
To determine the efficacy of the underlying protocols, network performance analysis is crucial during client-sensor interactions through the STGen middleware. Capturing real-time network traffic at the designated STGen middleware ports is achieved using TShark, the command-line version of Wireshark. However, raw traffic data lacks structure and requires transformation for meaningful analysis, necessitating the identification of key performance metrics such as latency, jitter, packet loss, retransmissions, and protocol overhead. To address this, Logstash preprocesses and transforms the raw traffic data into a structured format in real time, extracting relevant insights before transmitting it to Elasticsearch, where the data is indexed and stored for efficient querying and analysis \cite{elastic_kibana_visualizations}.

The STGen core node captures this network traffic by extracting real-time data from specific UDP ports and outputting it in JSON format, making it easier to feed into Elasticsearch for further processing. Researchers evaluating communication protocols can leverage this indexed data to analyze protocol performance under various conditions. By integrating Elasticsearch with Kibana, researchers gain access to dynamic, real-time visualizations that enable effective benchmarking of network behaviors. Kibana’s dashboards, time-series charts, and heatmaps provide insights into latency trends, traffic congestion points, and packet size distributions across different sensor nodes. By tailoring these visualizations to emphasize specific network performance metrics, researchers can extract actionable insights from STGen’s network traffic, optimizing both protocol design and middleware efficiency. Figure \ref{elk} illustrates real-time delay statistics visualized in Kibana, showcasing the effectiveness of this TShark and ELK workflow in STGen’s testbed analysis.

\end{description}

\begin{figure*}[htbp] % for sub figures over two columns in 
\centering
\subfloat[The start-up time of the STGen platform for varying the number of sensor nodes from 1K to 6K  under different RAM configurations.]{\includegraphics[width=0.45\columnwidth]{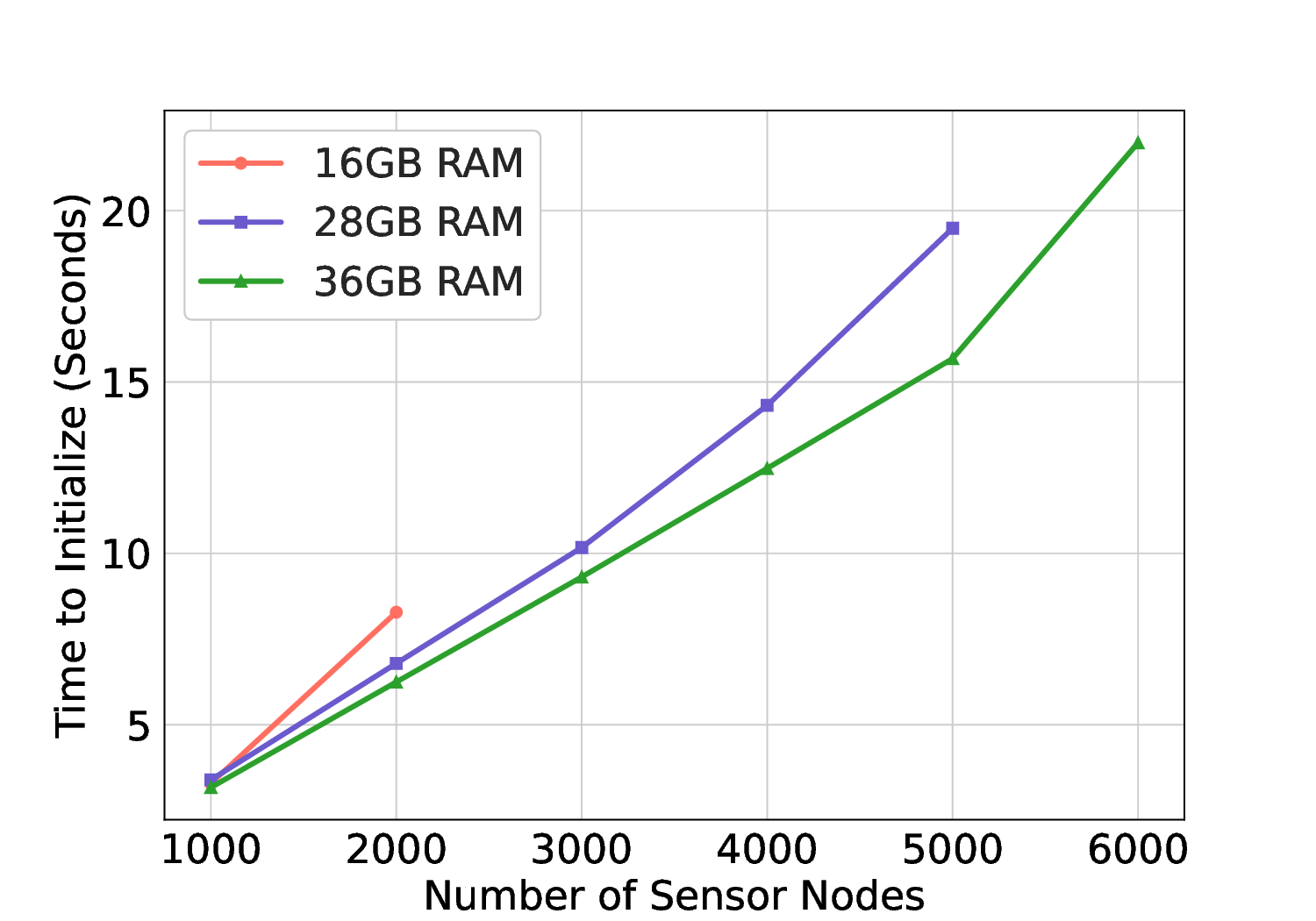} \label{fig:nodes_vs_time}}
\hfil
\subfloat[Memory footprint of the STGen platform for varying the number of sensor nodes from 1K to 6K  under different RAM configurations.]{\includegraphics[width=0.46\columnwidth]{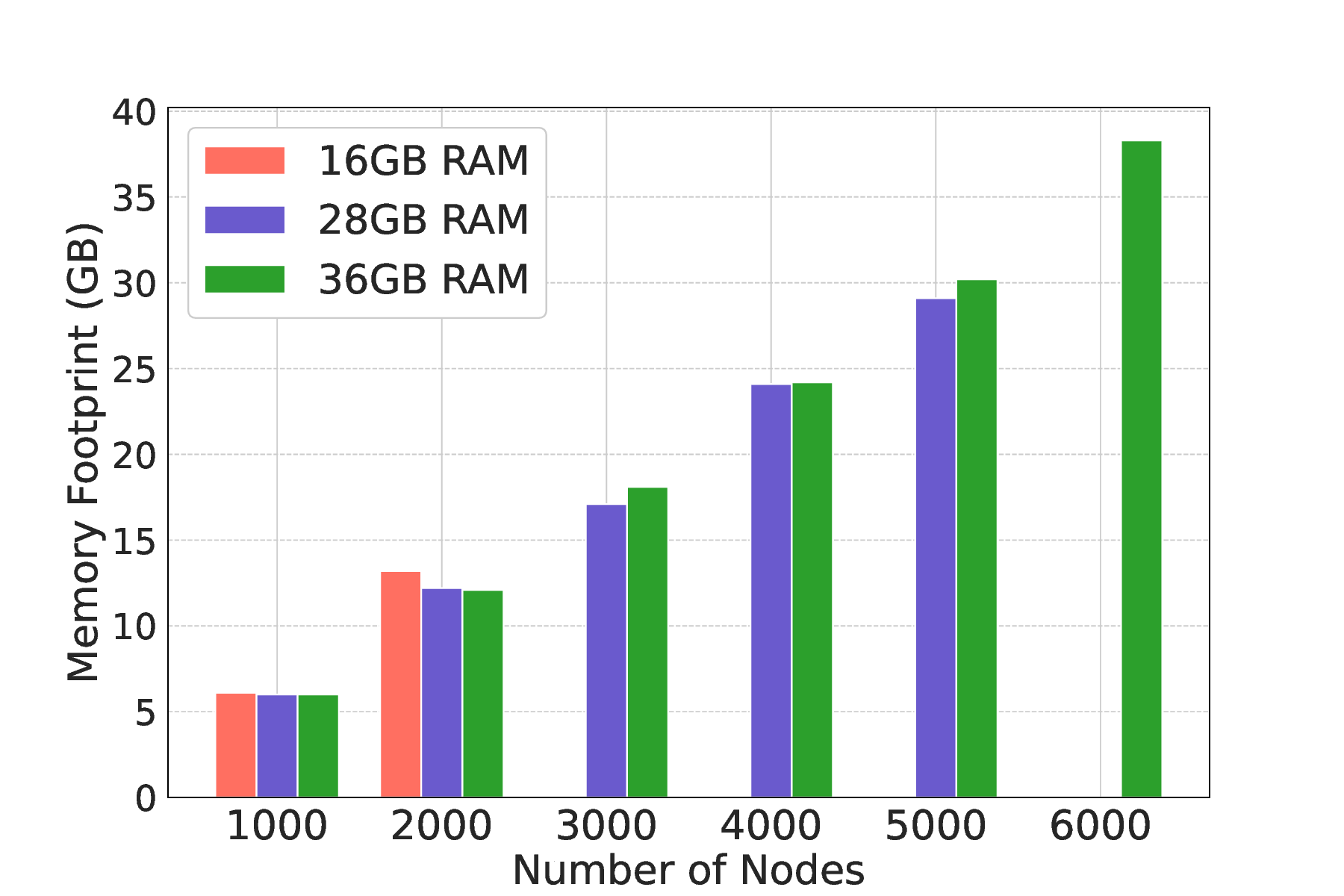} \label{fig:6k_nodes_memory}}
\caption{Performance Analysis of the STGen Platform Under Varying Sensor Node Configurations}
\label{fig:boot}
\end{figure*}

\begin{figure*}
\includegraphics[width=0.7\columnwidth]{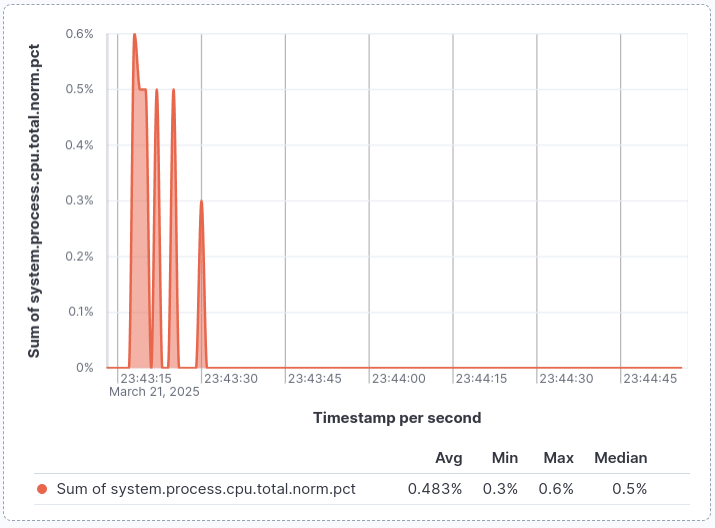} 
\caption{CPU (6 core) Utilization percentage of 3000 concurrent sensor node processes over a 5-minute monitoring interval.}
\label{fig:cpu_utilzation}
\end{figure*}

\section{Results and Observations}
\label{observations}

\begin{table*}[htbp]
    \centering
    \caption{Delay Statistics for Temperature Data Packet Transmission from STGen Core to Client at Varying Bandwidth and Packet Loss Rates.}
    \begin{tabular}{cccccc}
    
     \toprule
     Client & (Bandwidth, Packet Loss Rate) & Mean Delay (s) & Max Delay (s)& Min Delay (s)& Variance Delay (s)\\
     \midrule
        1 & (Unbounded, 0\%)   & 0.000224 & 0.001 & 0 & 0.0000001915 \\
        2 & (100 kbps, 5\%)   & 0.01297 & 0.013 & 0.012 & 0.00000049 \\
        3 & (100 kbps, 10\%) & 0.01228 & 0.013 & 0.012 & 0.000000483 \\
        4 & (10 kbps, 5\%)  & 0.07277 & 0.517 & 0.033 & 0.02515 \\
        5 & (10 kbps, 10\%)  & 0.04164 & 0.316000 & 0.034000 & 0.00191 \\
     \bottomrule
    \end{tabular}
    \label{tab:client_performance}
\end{table*}

\begin{figure*}[htbp] % for sub figures over two columns in 
\centering
\subfloat[Contribution of each sensors over 1 hour time intervals (1 hour)]{\includegraphics[width=0.49\columnwidth]{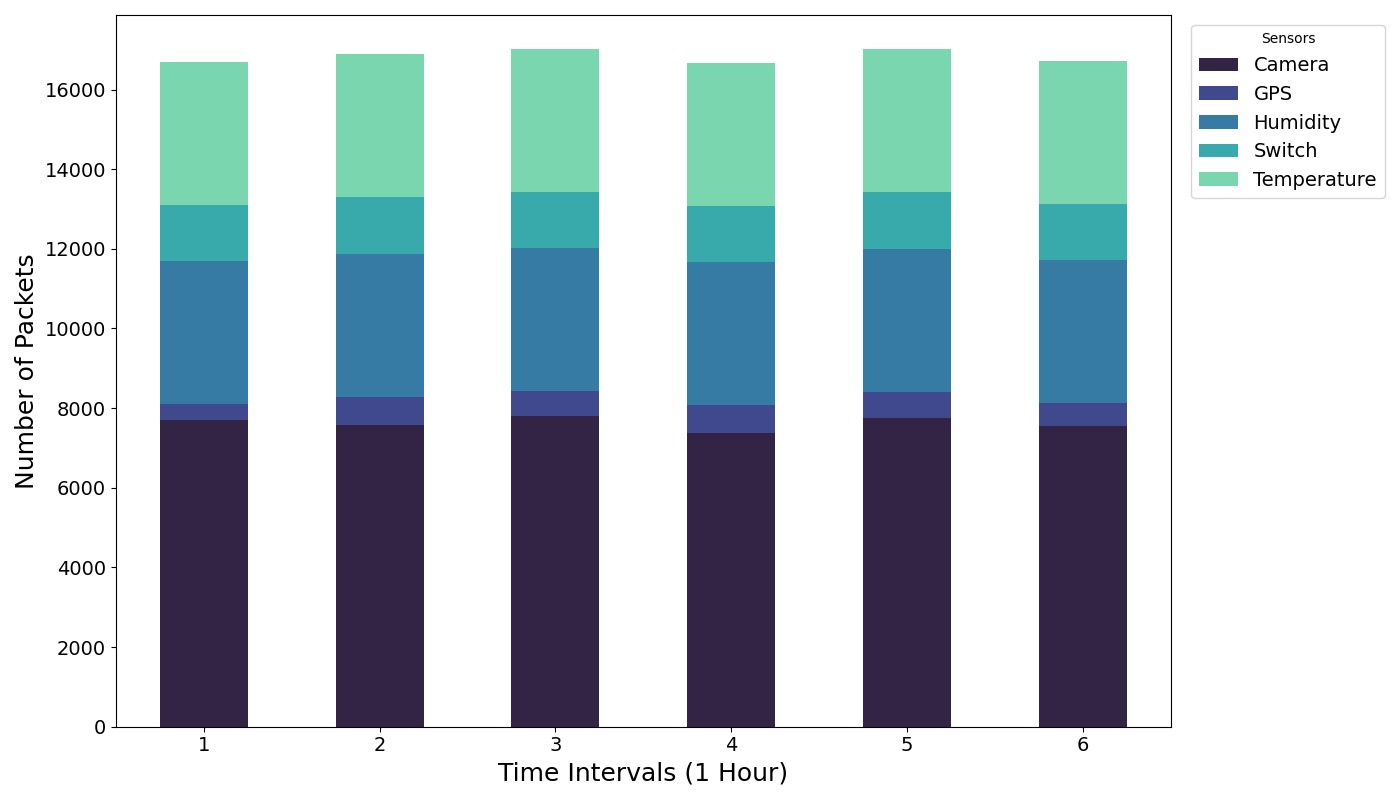} \label{fig:10a}}
\hfil
\subfloat[Frequency of each sensors]{\includegraphics[width=0.48\columnwidth]{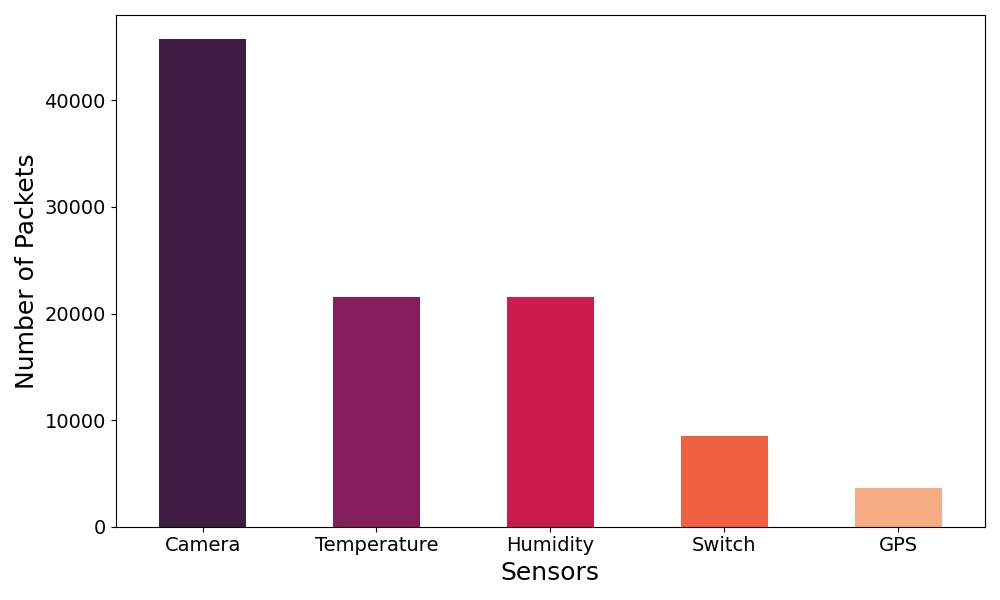} \label{fig:10b}}
\caption{Packet Distribution Analysis of different sensors deployed on the STGen Testbed.}
\label{fig:ten}
\end{figure*}

In this work, we have investigated the time taken to start up the STGen platform by varying the number of sensor nodes. In addition, the average latency for retrieving particular information from the STGen ecosystem has been considered an important KPI. The observations for the STGen system to boot are summarized as follows:

\begin{enumerate}
    \item Time taken to boot STGen platform consisting of six thousand sensors: 21.981 seconds
    \item Average latency to retrieve information from a particular sensor: 0.01965 seconds 
    \item Memory consumption by the STGen platform with the 6K nodes (including swap memory): 38.3 GB 
\end{enumerate}

Since the sensor nodes of the STGen are predominantly memory-constrained, the simulations were primarily conducted on a machine with 36 GB of RAM, with additional tests performed on machines with 16 GB and 28 GB of RAM. Each experiment was performed for 30 seconds while varying the number of nodes and maintaining an equal distribution of each sensor type. This procedure was carried out ten times to guarantee predictable outcomes. We observe from Figure \ref{fig:nodes_vs_time} \& \ref{fig:6k_nodes_memory} that decreasing memory not only limits the number of deployable nodes it also affects boot times with available memory. The root cause behind this observation may be the operating system, which spends extra time managing memory allocations.

To evaluate the system's capacity limits, all three components of the STGen core, client, and sensor nodes were deliberately co-located on a single machine. This configuration caused system breakdowns during the experiments due to substantial computational overhead and the simultaneous logging of sensor traffic data to local storage.

\begin{figure*} % for sub figures over two columns in 
\centering
\subfloat[Heatmap representation of temperature variations over time.]{\includegraphics[width=0.45\columnwidth]{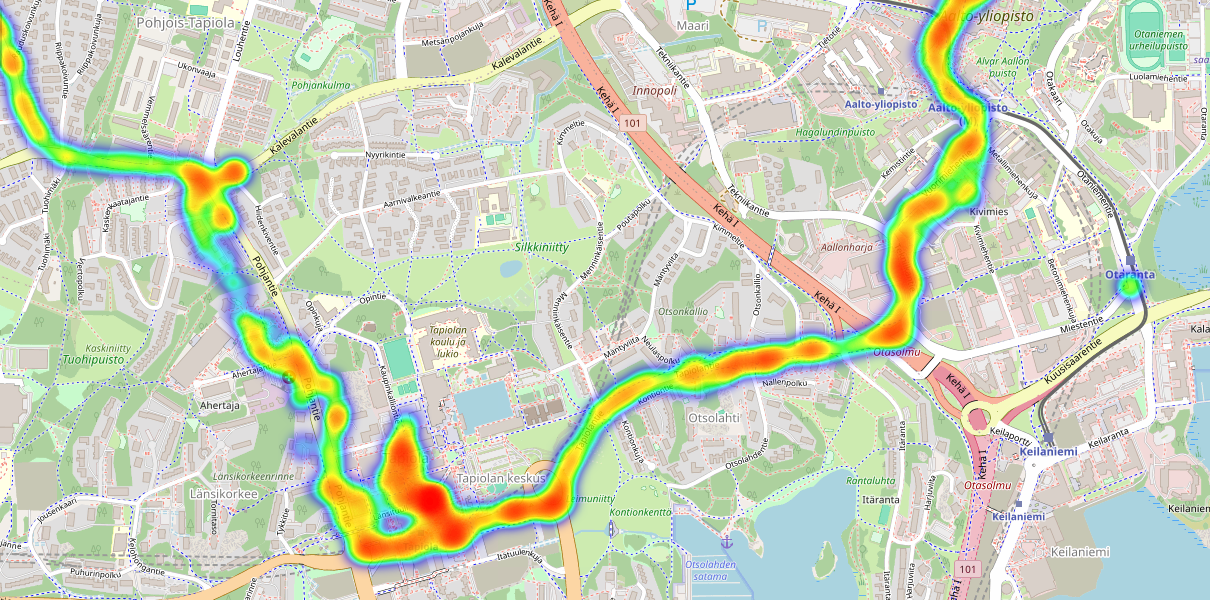} \label{fig:9a}}
\hfil
\subfloat[ON/OFF state visualization of switch activity along the route.]{\includegraphics[width=0.45\columnwidth]{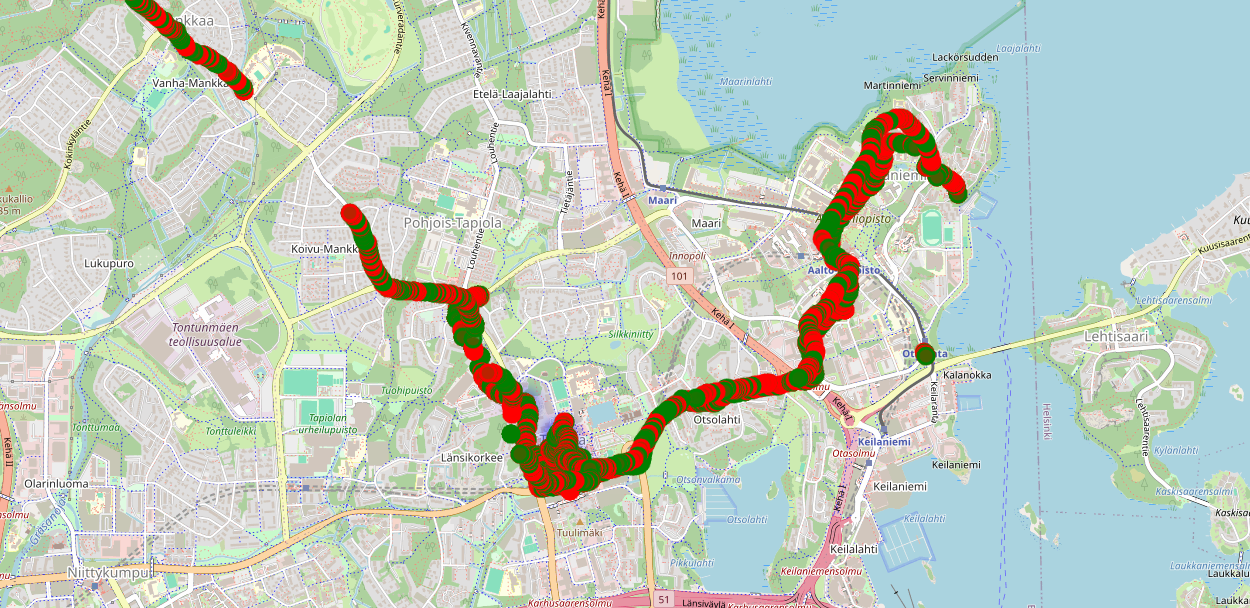} \label{fig:9b}}
\hfil
\subfloat[Spatial distribution of camera data, with intensity indicating high packet activity.]{\includegraphics[width=0.45\columnwidth]{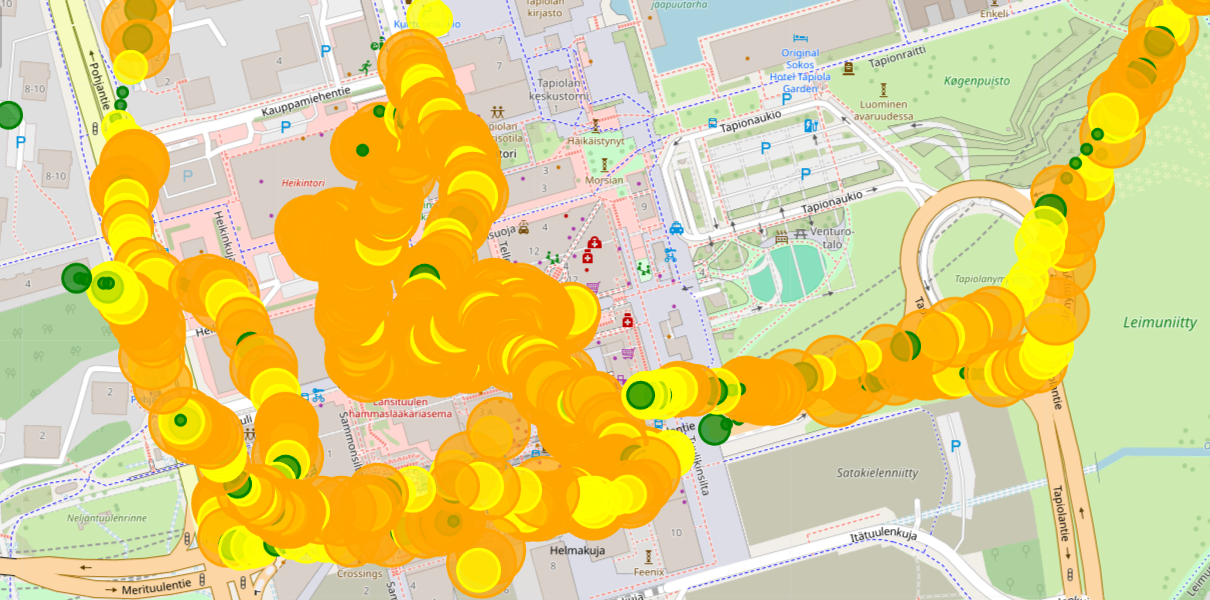} \label{fig:9c}}
\hfil
\subfloat[Visualizations of GPS sensor locations data.]{\includegraphics[width=0.45\columnwidth]{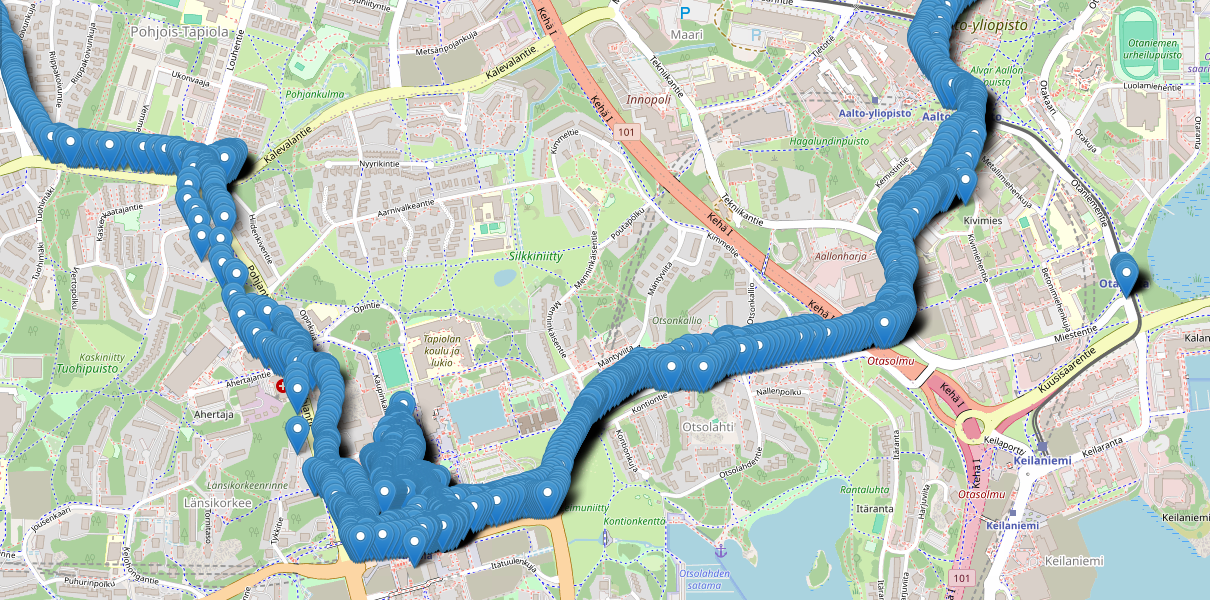} \label{fig:9d}}
\caption{Visualizations of packet transmission patterns over a geographic region, showcasing density, state changes, spatial distribution, and sensor interactions.}
\label{fig:9}
\end{figure*}

The experiment of benchmarking the startup of 7k sensor nodes has failed due to our experiment's limited system resources. We observe that initializing more than 6k sensor nodes causes issues with our experimental setup's available RAM, swap, and virtual memory. However, our workstation can handle up to 6k nodes with a memory footprint of about 38.3 GB, while the physical memory is only 36 GB. The benchmark is run with CLI commands, as the backend (Java) and frontend (Next.js) of the web application cause extra memory overhead. 
The \texttt{system.process.cpu.total.norm.pct} metric, as shown in Figure \ref{fig:cpu_utilzation}, represents the total percentage of CPU time consumed by 3000 sensor node processes. It was observed to peak at 0.6\% during sensor startup and remain at 0\% during subsequent periods when simulating thousands of sensor nodes concurrently. This behavior indicates minimal CPU usage per process, particularly during periods of low system load. Since the modules are decoupled, and in real-world scenarios, clients can request sensor data at any time, each module operates with an independent simulation period, ensuring they do not interfere with one another. The execution time \( T \) for generating sensor data is determined by the simulation time period \( S \):

\begin{equation}
T = S
\end{equation}

Every sensor type on the STGen testbed is programmed with the default transmission rate. The waiting time between consecutive data transmissions, which defines the data rate, can be adjusted without altering the execution time \( T \). Given an initial transmission interval \( I \) (the default timeout for a sensor type), the adjusted interval \( I' \) is determined by a percentage factor \( P \) (ranging from 1 to 100), which will be taken as input through the CLI arguments:

\begin{equation}
I' = I \times \frac{100}{P}
\end{equation}

An increase in \( P \) results in a decrease in \( I' \), leading to a higher data transmission rate. However, a decrease in \( P \) increases \( I' \), reducing the transmission frequency of the data. Since the execution time \( T \) remains unchanged, this approach allows dynamic control over data rates without interfering with the real-time execution of the sensor modules.

All simulations were conducted on 5 different components, such as temperature, humidity, GPS, switch, and camera, to ensure systematic analysis. Figure \ref{fig:10a} represents the number of packets transmitted over one-hour time intervals in a six-hour-long simulation. %over different one-hour time intervals.%
 The packets are categorized according to the contributing components: camera, humidity,  temperature, device, and GPS. The total number of packets remains relatively stable across the time intervals, with some variations in each sensor type's contribution, which is shown in Figure \ref{fig:10b}. The camera sensors are the most data-intensive due to high-resolution data transmission, whereas devices and GPS sensors contribute significantly less. 

Figure \ref{fig:9} is derived from the simulated dataset for elaborated visual representations, providing insights into sensor activities and their spatial-temporal patterns. These visualizations are crucial for understanding the data quality and behavior of simulated sensor nodes, helping improve data transmission efficiency, and anomaly detection. Figure \ref{fig:9a} is a heatmap of temperature fluctuations along the simulated path, where warmer colors represent higher temperatures and lighter colors represent lower temperatures. Similarly, A binary visualization of switch ON/OFF states along the route is depicted in Figure \ref{fig:9b}, with active states in green and inactive (OFF) states in red. The visualization of Figure \ref{fig:9c} highlights regions with frequent image or video capture events. Here, the intensity of yellow and orange hues signifies higher packer activity. Lastly, Figure \ref{fig:9d} provides the correlation between GPS sensor locations and packet transmission counts. These insights, generated from STGen simulations, enable researchers to assess and optimize the quality of traffic data generated by simulated sensor nodes.

\lstset{
    basicstyle=\ttfamily\small,
    frame=single,
    breaklines=true,
    breakatwhitespace=false, 
    postbreak=\mbox{$\hookrightarrow$\space}
}

\begin{figure}[h!]
% \centering
% % \begin{lstlisting}
% % watch -n 1 "
% % total=0; \
% % for process in python STGen; do \
% %     for pid in \$(pgrep \$process); do \
% %         total=\$((total + \$(pmap \$pid | grep total | awk '{print \$2}' | sed 's/K//'))); \
% %     done; \
% % done; \
% % echo \"Total Memory: \$((total / (1024 * 1024))) GB\""
% % \end{lstlisting}
% \caption{Bash script to monitor memory usage of Python and STGen processes in GB.}
% \label{tab:pscommand}

\end{figure}

\lstset{
    basicstyle=\ttfamily\small,
    frame=single,
    breaklines=true,
    breakatwhitespace=false, 
    postbreak=\mbox{$\hookrightarrow$\space}
}

\section{Comparision with Gotham \& GothX}
\label{discussion}

\begin{figure}[h]
  \centering
    \includegraphics[width=0.7\textwidth]{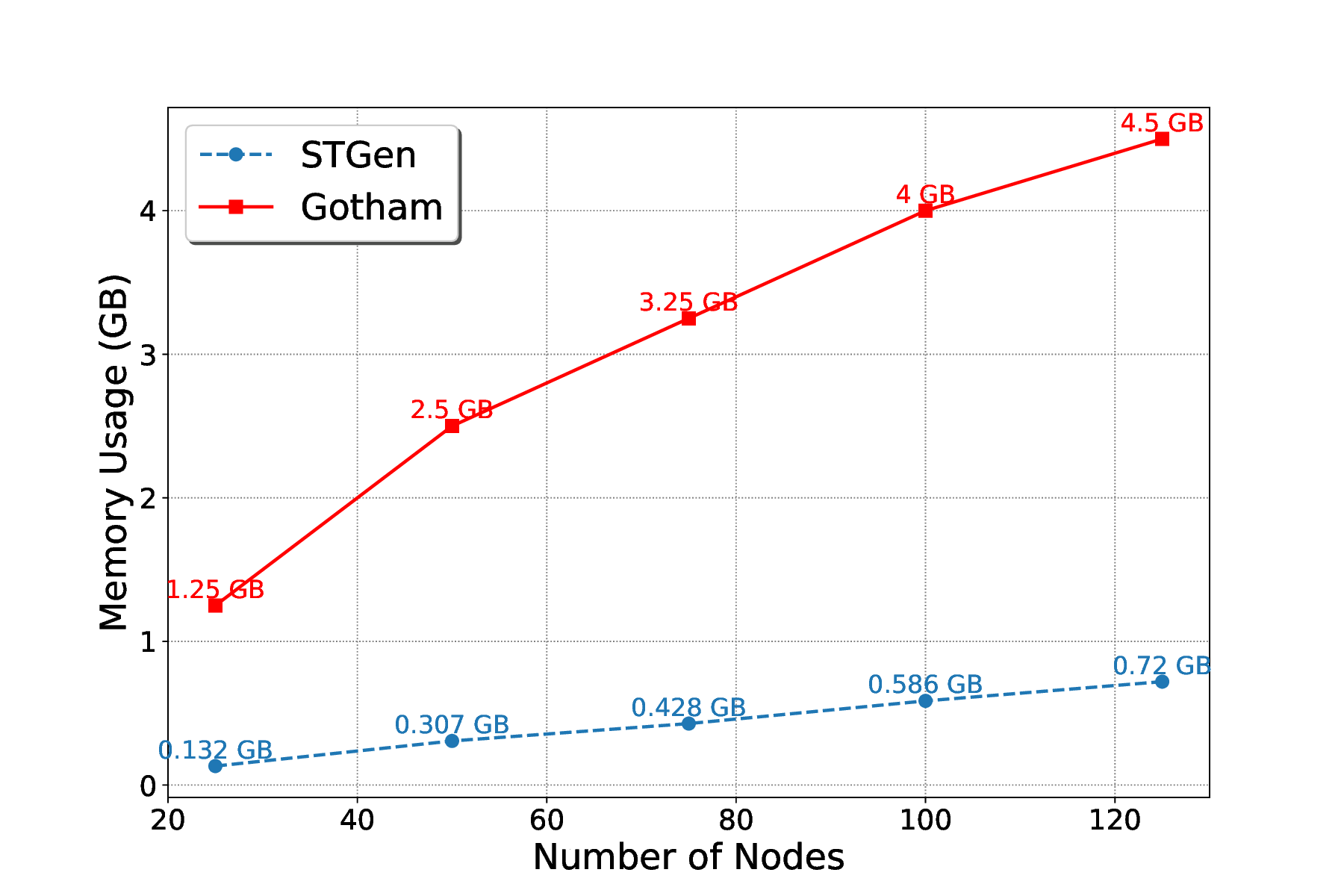}
  \caption{Memory Footprint Comparison: STGen vs Gotham for Varying Sensor Node Counts. }
  \label{stgvsgotham}
\end{figure}

In this article, STGen has been compared with the two most popular testbeds, Gotham and GothX, respectively, with the latter being developed as an improved version of the former to meet further demands and challenges in IoT simulations. The purpose of this comparison is to demonstrate the effectiveness of STGen in terms of resource utilization and startup time in performing large-scale simulations. As shown in Figure \ref{stgvsgotham}, STGen consistently outperforms Gotham in terms of memory usage, with reductions ranging between 84\% and 89\% as the number of nodes increases. The percentage reduction decreases slightly as the number of nodes increases. This result indicates a diminishing return but still a substantial reduction overall. The lightweight design of STGen results in approximately 87.5\% lower memory usage, requiring just 2.56 GB of RAM to initialize 450 nodes, compared to the 20.4 GB exploited by GothX. GothX takes approximately 26 minutes to set up a large topology with four VM nodes and 498 Docker nodes, whereas STGen takes just 1.645 seconds to initialize 500 sensor nodes along with one client and one core node. This represents a significant improvement, reducing the setup time by 99.9\%. Despite the fact that virtual machines (VMs) and Docker nodes are effective in complex virtualized environments, it is clear that Gotham and GothX incur significant overhead due to their reliance on them. The lightweight architecture of STGen thus provides a more efficient and rapid initialization process, making it highly suitable for scenarios requiring quick setup and scalable IoT simulations.

\begin{figure}[!t]
  \centering  \includegraphics[width=.98\textwidth]{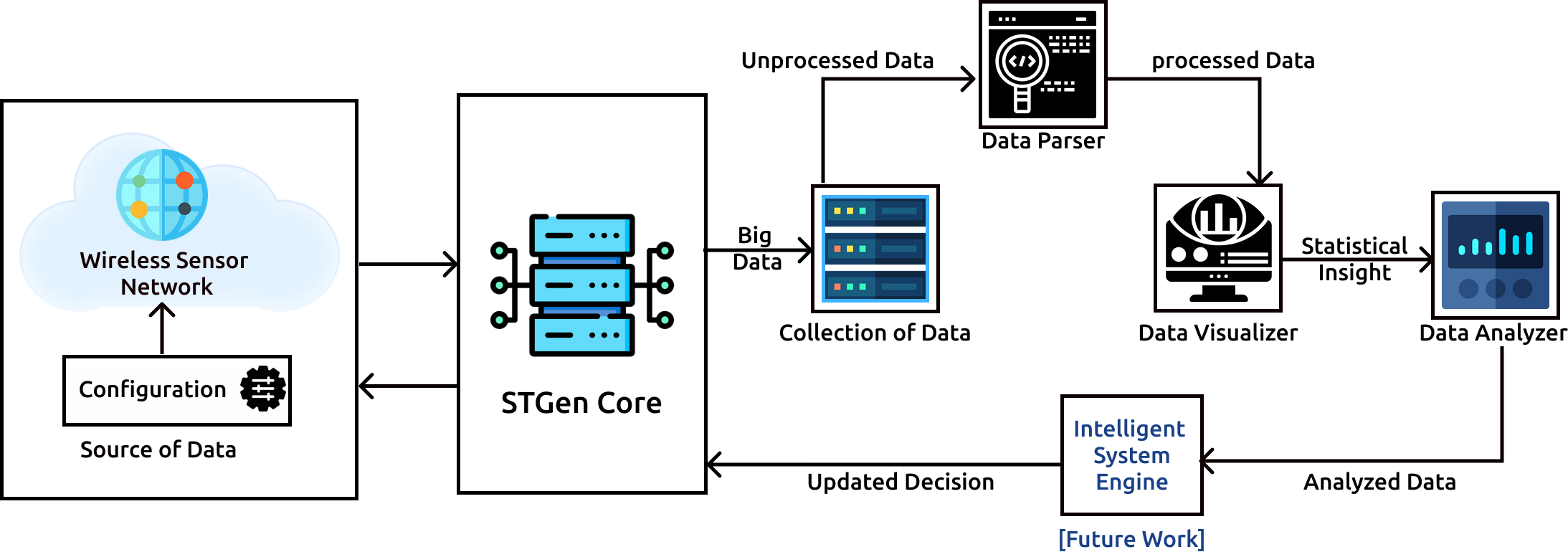}
  \caption{The Ecosystem of the distributed STGen Testbed and potential future work based on Big Data Analytics }
  \label{stgen-futurework}
\end{figure}

Gotham \& GothX enables the deployment of complex network topologies by leveraging virtual machine (VM)-based nodes, each requiring substantial computational resources. However, this approach incurs significant processing overhead, which makes large-scale IoT simulations on a single machine computationally impractical. The allocation of resources to multiple VM nodes leads to excessive CPU and memory consumption, resulting in system bottlenecks and resource contention. Consequently, the high computational demand restricts scalability and degrades the performance of other concurrent processes running on the system. To maintain a lightweight architecture, STGen does not natively support the deployment of complex network topologies; rather, it prioritizes a streamlined setup process to facilitate efficient protocol evaluation and experimentation.

The features of STGen enhance its capability and scalability as a tool for the experimentation of the IoT protocol while ensuring adaptability for future developments. STGen provides a solid basis for simulations of varying scales, effectively connecting controlled testing settings. In addition, STGen supports comprehensive guidance on streamlining the process of test environment setup and maintenance to facilitate seamless IoT testing composed of different types of sensors. The Web-Based UI adds easy interaction and visualization of IoT simulations in the user-friendly web application. The modular architecture of STGen allows for the distributed deployment of the entire ecosystem across various premises. This feature enables STGen to scale from thousands to millions of sensor nodes with almost no configuration effort.

\section{Future Work}
\label{future}

This section explains how STGen can serve as a suitable tool for research on the Internet of Things (IoT) to automate decision-making and Big Data Analytics. According to the figure \ref{stgen-futurework}, The STGen core, acting as a middleware, has the potential to serve as a large-scale data generator. The generated data, after being parsed into the processed data, can be fed into a statistical tool to derive meaningful statistical insights. To ensure that the STGen testbed remains competitive and future-ready, its development will be guided by many crucial areas. Traffic data will initially undergo preprocessing, followed by comprehensive visualization and analysis utilizing machine learning methods. Based on analysis, the intelligent system can dynamically adjust the STGen core parameters to make traffic more realistic and contextually relevant. Since all traffic data and network packets will be analyzed, predictive analytics will be used to proactively detect system performance bottlenecks and optimize data generation processes in real time. This adaptive approach will improve automation and efficiency while refining intelligence in IoT simulations, paving the way for more responsive and self-optimizing systems. However, considering the low-cost model of cloud resources, usability and accessibility can be further enhanced by deploying the STGen testbed in the cloud using Terraform (Infrastructure as Code) \cite{hashicorp_aws_get_started} for large-scale testbed simulations.

\section{Conclusion}
\label{conclusion}

This paper introduced STGen, a scalable and cost-effective sensor traffic generator designed to facilitate the experimentation of IoT protocols by emulating wireless sensor networks (WSNs) in a hybrid environment that addresses key challenges associated with the deployment of real-world testbeds, including financial constraints, resource limitations, and scalability problems, by providing a lightweight yet powerful alternative. Its modular design enables researchers to conduct empirical studies on IoT protocols, which also allow new integrations or enhancements to be developed and deployed in isolation without affecting other modules, such as the integration of the ELK stack, enabling real-time traffic analysis, offering immediate insights into sensor activity and protocol behavior. Researchers can interact with STGen through various interfaces, a command-line tool (CLI), a web dashboard, and REST API endpoints following OpenAPI standards. Researchers can utilize these interfaces for a wide range of experiments, such as simulating and mitigating DoS attacks, automating test environments, and assessing protocol performance under high-traffic loads. Although most popular testbeds struggle to run hundreds of sensor nodes on a single machine, the lightweight design of STGen enables it to handle thousands with ease. STGen offers a fast start time of 21.981 s for 6,000 nodes and uses only 38.3 GB of memory, outperforming recent testbeds like Gotham and GothX. Looking ahead, a key direction for STGen is automating the deployment in cloud environments using Terraform, an Infrastructure-as-Code (IaC) tool that ensures scalable, repeatable, and automated provisioning of cloud resources.

% \bibliographystyle{ACM-Reference-Format}


\begin{thebibliography}{31}

%%% the Web address.

\ifx \showCODEN    \undefined \def \showCODEN     #1{\unskip}     \fi
\ifx \showISBNx    \undefined \def \showISBNx     #1{\unskip}     \fi
\ifx \showISBNxiii \undefined \def \showISBNxiii  #1{\unskip}     \fi
\ifx \showISSN     \undefined \def \showISSN      #1{\unskip}     \fi
\ifx \showLCCN     \undefined \def \showLCCN      #1{\unskip}     \fi
\ifx \shownote     \undefined \def \shownote      #1{#1}          \fi
\ifx \showarticletitle \undefined \def \showarticletitle #1{#1}   \fi
\ifx \showURL      \undefined \def \showURL       {\relax}        \fi
% invisible to TeX
\providecommand\bibfield[2]{#2}
\providecommand\bibinfo[2]{#2}
\providecommand\natexlab[1]{#1}
\providecommand\showeprint[2][]{arXiv:#2}

\bibitem[Adjih et~al\mbox{.}(2015)]%
        {adjih2015fit}
\bibfield{author}{\bibinfo{person}{Cedric Adjih}, \bibinfo{person}{Emmanuel
  Baccelli}, \bibinfo{person}{Eric Fleury}, \bibinfo{person}{Gaetan Harter},
  \bibinfo{person}{Nathalie Mitton}, \bibinfo{person}{Thomas Noel},
  \bibinfo{person}{Roger Pissard-Gibollet}, \bibinfo{person}{Frederic
  Saint-Marcel}, \bibinfo{person}{Guillaume Schreiner}, \bibinfo{person}{Julien
  Vandaele}, {et~al\mbox{.}}} \bibinfo{year}{2015}\natexlab{}.
\newblock \showarticletitle{FIT IoT-LAB: A large scale open experimental IoT
  testbed}. In \bibinfo{booktitle}{\emph{2015 IEEE 2nd World Forum on Internet
  of Things (WF-IoT)}}. IEEE, \bibinfo{pages}{459--464}.
\newblock


\bibitem[Almutairi et~al\mbox{.}(2024)]%
        {almutairi2024advancements}
\bibfield{author}{\bibinfo{person}{Reham Almutairi}, \bibinfo{person}{Giacomo
  Bergami}, {and} \bibinfo{person}{Graham Morgan}.}
  \bibinfo{year}{2024}\natexlab{}.
\newblock \showarticletitle{Advancements and challenges in IoT simulators: A
  comprehensive review}.
\newblock \bibinfo{journal}{\emph{Sensors}} \bibinfo{volume}{24},
  \bibinfo{number}{5} (\bibinfo{year}{2024}), \bibinfo{pages}{1511}.
\newblock
\urldef\tempurl%
\url{https://doi.org/10.3390/s24051511}
\showURL{%
\tempurl}


\bibitem[Alsukayti(2020)]%
        {alsukayti2020multidimensional}
\bibfield{author}{\bibinfo{person}{Ibrahim~S Alsukayti}.}
  \bibinfo{year}{2020}\natexlab{}.
\newblock \showarticletitle{A multidimensional internet of things testbed
  system: Development and evaluation}.
\newblock \bibinfo{journal}{\emph{Wireless Communications and Mobile
  Computing}} \bibinfo{volume}{2020}, \bibinfo{number}{1}
  (\bibinfo{year}{2020}), \bibinfo{pages}{8849433}.
\newblock
\urldef\tempurl%
\url{https://doi.org/10.1155/2020/8849433}
\showURL{%
\tempurl}


\bibitem[Buratti et~al\mbox{.}(2015)]%
        {buratti2015testing}
\bibfield{author}{\bibinfo{person}{Chiara Buratti}, \bibinfo{person}{Andrea
  Stajkic}, \bibinfo{person}{Gordana Gardasevic}, \bibinfo{person}{Sebastiano
  Milardo}, \bibinfo{person}{M~Danilo Abrignani}, \bibinfo{person}{Stefan
  Mijovic}, \bibinfo{person}{Giacomo Morabito}, {and} \bibinfo{person}{Roberto
  Verdone}.} \bibinfo{year}{2015}\natexlab{}.
\newblock \showarticletitle{Testing protocols for the internet of things on the
  EuWIn platform}.
\newblock \bibinfo{journal}{\emph{IEEE Internet of Things Journal}}
  \bibinfo{volume}{3}, \bibinfo{number}{1} (\bibinfo{year}{2015}),
  \bibinfo{pages}{124--133}.
\newblock
\urldef\tempurl%
\url{https://doi.org/10.1109/jiot.2015.2462030}
\showURL{%
\tempurl}


\bibitem[Bures et~al\mbox{.}(2021)]%
        {bures2021patriot}
\bibfield{author}{\bibinfo{person}{Miroslav Bures}, \bibinfo{person}{Bestoun~S
  Ahmed}, \bibinfo{person}{Vaclav Rechtberger}, \bibinfo{person}{Matej Klima},
  \bibinfo{person}{Michal Trnka}, \bibinfo{person}{Miroslav Jaros},
  \bibinfo{person}{Xavier Bellekens}, \bibinfo{person}{Dani Almog}, {and}
  \bibinfo{person}{Pavel Herout}.} \bibinfo{year}{2021}\natexlab{}.
\newblock \showarticletitle{Patriot: Iot automated interoperability and
  integration testing framework}. In \bibinfo{booktitle}{\emph{2021 14th IEEE
  Conference on Software Testing, Verification and Validation (ICST)}}. IEEE,
  \bibinfo{pages}{454--459}.
\newblock
\urldef\tempurl%
\url{https://doi.org/10.1109/icst49551.2021.00059}
\showURL{%
\tempurl}


\bibitem[Chernyshev et~al\mbox{.}(2017)]%
        {chernyshev2017internet}
\bibfield{author}{\bibinfo{person}{Maxim Chernyshev}, \bibinfo{person}{Zubair
  Baig}, \bibinfo{person}{Oladayo Bello}, {and} \bibinfo{person}{Sherali
  Zeadally}.} \bibinfo{year}{2017}\natexlab{}.
\newblock \showarticletitle{Internet of things (iot): Research, simulators, and
  testbeds}.
\newblock \bibinfo{journal}{\emph{IEEE Internet of Things Journal}}
  \bibinfo{volume}{5}, \bibinfo{number}{3} (\bibinfo{year}{2017}),
  \bibinfo{pages}{1637--1647}.
\newblock
\urldef\tempurl%
\url{https://doi.org/10.1109/jiot.2017.2786639}
\showURL{%
\tempurl}


\bibitem[Developers(2020)]%
        {ubuntu_tc}
\bibfield{author}{\bibinfo{person}{Ubuntu Developers}.}
  \bibinfo{year}{2020}\natexlab{}.
\newblock \bibinfo{title}{tc - Linux traffic control utility}.
\newblock
\urldef\tempurl%
\url{https://manpages.ubuntu.com/manpages/focal/en/man8/tc.8.html}
\showURL{%
Retrieved 2025-03-22 from \tempurl}
\newblock
\shownote{Accessed: March 22, 2025}.


\bibitem[Eddy(2022)]%
        {eddy2022rfc}
\bibfield{author}{\bibinfo{person}{W. Eddy}.} \bibinfo{year}{2022}\natexlab{}.
\newblock \bibinfo{title}{RFC 9293: Transmission Control Protocol (TCP)}.
\newblock
\urldef\tempurl%
\url{https://datatracker.ietf.org/doc/html/rfc9293}
\showURL{%
Retrieved 2025-03-23 from \tempurl}
\newblock
\shownote{RFC Editor}.


\bibitem[Eggert et~al\mbox{.}(2017)]%
        {eggert2017rfc}
\bibfield{author}{\bibinfo{person}{L. Eggert}, \bibinfo{person}{G. Fairhurst},
  {and} \bibinfo{person}{G. Shepherd}.} \bibinfo{year}{2017}\natexlab{}.
\newblock \bibinfo{title}{RFC 8085: UDP Usage Guidelines}.
\newblock
\urldef\tempurl%
\url{https://datatracker.ietf.org/doc/html/rfc8085}
\showURL{%
Retrieved 2025-03-23 from \tempurl}
\newblock
\shownote{RFC Editor}.


\bibitem[Elastic(nd)]%
        {elastic_kibana_visualizations}
\bibfield{author}{\bibinfo{person}{Elastic}.} \bibinfo{year}{n.d.}\natexlab{}.
\newblock \bibinfo{title}{Create visualizations in Kibana}.
\newblock
\urldef\tempurl%
\url{https://www.elastic.co/guide/en/kibana/6.8/createvis.html}
\showURL{%
\tempurl}
\newblock
\shownote{Accessed: 2025-03-20}.


\bibitem[Falkenberg et~al\mbox{.}(2017)]%
        {falkenberg2017phynetlab}
\bibfield{author}{\bibinfo{person}{Robert Falkenberg}, \bibinfo{person}{Mojtaba
  Masoudinejad}, \bibinfo{person}{Markus Buschhoff}, \bibinfo{person}{Aswin
  Karthik~Ramachandran Venkatapathy}, \bibinfo{person}{Birte Friesel},
  \bibinfo{person}{Michael ten Hompel}, \bibinfo{person}{Olaf Spinczyk}, {and}
  \bibinfo{person}{Christian Wietfeld}.} \bibinfo{year}{2017}\natexlab{}.
\newblock \showarticletitle{PhyNetLab: An IoT-based warehouse testbed}. In
  \bibinfo{booktitle}{\emph{2017 Federated Conference on Computer Science and
  Information Systems (FedCSIS)}}. IEEE, \bibinfo{pages}{1051--1055}.
\newblock
\urldef\tempurl%
\url{https://doi.org/10.15439/2017f267}
\showURL{%
\tempurl}


\bibitem[Ghazanfar et~al\mbox{.}(2020)]%
        {ghazanfar2020iot}
\bibfield{author}{\bibinfo{person}{Syed Ghazanfar}, \bibinfo{person}{Faisal
  Hussain}, \bibinfo{person}{Atiq~Ur Rehman}, \bibinfo{person}{Ubaid~U Fayyaz},
  \bibinfo{person}{Farrukh Shahzad}, {and} \bibinfo{person}{Ghalib~A Shah}.}
  \bibinfo{year}{2020}\natexlab{}.
\newblock \showarticletitle{Iot-flock: An open-source framework for iot traffic
  generation}. In \bibinfo{booktitle}{\emph{2020 International Conference on
  Emerging Trends in Smart Technologies (ICETST)}}. IEEE,
  \bibinfo{pages}{1--6}.
\newblock
\urldef\tempurl%
\url{https://doi.org/10.1109/icetst49965.2020.9080732}
\showURL{%
\tempurl}


\bibitem[HashiCorp(nd)]%
        {hashicorp_aws_get_started}
\bibfield{author}{\bibinfo{person}{HashiCorp}.}
  \bibinfo{year}{n.d.}\natexlab{}.
\newblock \bibinfo{title}{Get Started with AWS Using Terraform}.
\newblock
\urldef\tempurl%
\url{https://developer.hashicorp.com/terraform/tutorials/aws-get-started/aws-build}
\showURL{%
Retrieved 2025-02-23 from \tempurl}
\newblock
\shownote{Accessed: March 20, 2025}.


\bibitem[Jiang et~al\mbox{.}(2020)]%
        {jiang2020makesense}
\bibfield{author}{\bibinfo{person}{Jie Jiang}, \bibinfo{person}{Riccardo
  Pozza}, \bibinfo{person}{Nigel Gilbert}, {and} \bibinfo{person}{Klaus
  Moessner}.} \bibinfo{year}{2020}\natexlab{}.
\newblock \showarticletitle{Makesense: An iot testbed for social research of
  indoor activities}.
\newblock \bibinfo{journal}{\emph{ACM Transactions on Internet of Things}}
  \bibinfo{volume}{1}, \bibinfo{number}{3} (\bibinfo{year}{2020}),
  \bibinfo{pages}{1--25}.
\newblock
\urldef\tempurl%
\url{https://doi.org/10.1145/3381914}
\showURL{%
\tempurl}


\bibitem[Minani et~al\mbox{.}(2024)]%
        {minani2024systematic}
\bibfield{author}{\bibinfo{person}{Jean~Baptiste Minani},
  \bibinfo{person}{Fatima Sabir}, \bibinfo{person}{Naouel Moha}, {and}
  \bibinfo{person}{Yann-Ga{\"e}l Gu{\'e}h{\'e}neuc}.}
  \bibinfo{year}{2024}\natexlab{}.
\newblock \showarticletitle{A systematic review of IoT systems testing:
  Objectives, approaches, tools, and challenges}.
\newblock \bibinfo{journal}{\emph{IEEE Transactions on Software Engineering}}
  (\bibinfo{year}{2024}), \bibinfo{pages}{785--815}.
\newblock
\urldef\tempurl%
\url{https://doi.org/10.1109/TSE.2024.3363611}
\showURL{%
\tempurl}


\bibitem[Moon et~al\mbox{.}(2019)]%
        {moon2019heterogeneous}
\bibfield{author}{\bibinfo{person}{Jaewon Moon}, \bibinfo{person}{Seungwoo
  Kum}, {and} \bibinfo{person}{Sangwon Lee}.} \bibinfo{year}{2019}\natexlab{}.
\newblock \showarticletitle{A heterogeneous IoT data analysis framework with
  collaboration of edge-cloud computing: Focusing on indoor PM10 and PM2. 5
  status prediction}.
\newblock \bibinfo{journal}{\emph{Sensors}} \bibinfo{volume}{19},
  \bibinfo{number}{14} (\bibinfo{year}{2019}), \bibinfo{pages}{3038}.
\newblock
\urldef\tempurl%
\url{https://doi.org/10.3390/s19143038}
\showURL{%
\tempurl}


\bibitem[Munoz et~al\mbox{.}(2019)]%
        {munoz2019opentestbed}
\bibfield{author}{\bibinfo{person}{Jonathan Munoz}, \bibinfo{person}{Fabian
  Rincon}, \bibinfo{person}{Tengfei Chang}, \bibinfo{person}{Xavier
  Vilajosana}, \bibinfo{person}{Brecht Vermeulen}, \bibinfo{person}{Thijs
  Walcarius}, \bibinfo{person}{Wim Van~de Meerssche}, {and}
  \bibinfo{person}{Thomas Watteyne}.} \bibinfo{year}{2019}\natexlab{}.
\newblock \showarticletitle{OpenTestBed: Poor man's IoT testbed}. In
  \bibinfo{booktitle}{\emph{IEEE INFOCOM 2019-IEEE Conference on Computer
  Communications Workshops (INFOCOM WKSHPS)}}. IEEE, \bibinfo{pages}{467--471}.
\newblock
\urldef\tempurl%
\url{https://doi.org/10.1109/INFCOMW.2019.8845269}
\showURL{%
\tempurl}


\bibitem[Ngo et~al\mbox{.}(2023)]%
        {ngo2023new}
\bibfield{author}{\bibinfo{person}{Thi Thu~Trang Ngo}, \bibinfo{person}{David
  Sarramia}, \bibinfo{person}{Myoung-Ah Kang}, {and}
  \bibinfo{person}{Fran{\c{c}}ois Pinet}.} \bibinfo{year}{2023}\natexlab{}.
\newblock \showarticletitle{A new approach based on ELK stack for the analysis
  and visualisation of geo-referenced sensor data}.
\newblock \bibinfo{journal}{\emph{SN Computer Science}} \bibinfo{volume}{4},
  \bibinfo{number}{3} (\bibinfo{year}{2023}), \bibinfo{pages}{241}.
\newblock
\urldef\tempurl%
\url{https://doi.org/10.1007/s42979-022-01628-6}
\showURL{%
\tempurl}


\bibitem[Noaman et~al\mbox{.}(2022)]%
        {noaman2022challenges}
\bibfield{author}{\bibinfo{person}{Muhammad Noaman},
  \bibinfo{person}{Muhammad~Sohail Khan}, \bibinfo{person}{Muhammad~Faisal
  Abrar}, \bibinfo{person}{Sikandar Ali}, \bibinfo{person}{Atif Alvi}, {and}
  \bibinfo{person}{Muhammad~Asif Saleem}.} \bibinfo{year}{2022}\natexlab{}.
\newblock \showarticletitle{Challenges in integration of heterogeneous internet
  of things}.
\newblock \bibinfo{journal}{\emph{Scientific Programming}}
  \bibinfo{volume}{2022}, \bibinfo{number}{1} (\bibinfo{year}{2022}),
  \bibinfo{pages}{1–--14}.
\newblock
\urldef\tempurl%
\url{https://doi.org/10.1155/2022/8626882}
\showURL{%
\tempurl}


\bibitem[Pan et~al\mbox{.}(2015)]%
        {pan2015internet}
\bibfield{author}{\bibinfo{person}{Jianli Pan}, \bibinfo{person}{Raj Jain},
  \bibinfo{person}{Subharthi Paul}, \bibinfo{person}{Tam Vu},
  \bibinfo{person}{Abusayeed Saifullah}, {and} \bibinfo{person}{Mo Sha}.}
  \bibinfo{year}{2015}\natexlab{}.
\newblock \showarticletitle{An internet of things framework for smart energy in
  buildings: designs, prototype, and experiments}.
\newblock \bibinfo{journal}{\emph{IEEE internet of things journal}}
  \bibinfo{volume}{2}, \bibinfo{number}{6} (\bibinfo{year}{2015}),
  \bibinfo{pages}{527--537}.
\newblock
\urldef\tempurl%
\url{https://doi.org/10.1109/jiot.2015.2413397}
\showURL{%
\tempurl}


\bibitem[Papadopoulos et~al\mbox{.}(2013)]%
        {papadopoulos2013adding}
\bibfield{author}{\bibinfo{person}{Georgios~Z Papadopoulos},
  \bibinfo{person}{Julien Beaudaux}, \bibinfo{person}{Antoine Gallais},
  \bibinfo{person}{Thomas Noel}, {and} \bibinfo{person}{Guillaume Schreiner}.}
  \bibinfo{year}{2013}\natexlab{}.
\newblock \showarticletitle{Adding value to WSN simulation using the IoT-LAB
  experimental platform}. In \bibinfo{booktitle}{\emph{2013 IEEE 9th
  International Conference on Wireless and Mobile Computing, Networking and
  Communications (WiMob)}}. IEEE, \bibinfo{pages}{485--490}.
\newblock
\urldef\tempurl%
\url{https://doi.org/10.1109/WiMOB.2013.6673403}
\showURL{%
\tempurl}


\bibitem[Papadopoulos et~al\mbox{.}(2017)]%
        {papadopoulos2017thorough}
\bibfield{author}{\bibinfo{person}{Georgios~Z Papadopoulos},
  \bibinfo{person}{Antoine Gallais}, \bibinfo{person}{Guillaume Schreiner},
  \bibinfo{person}{Emery Jou}, {and} \bibinfo{person}{Thomas Noel}.}
  \bibinfo{year}{2017}\natexlab{}.
\newblock \showarticletitle{Thorough IoT testbed characterization: From
  proof-of-concept to repeatable experimentations}.
\newblock \bibinfo{journal}{\emph{Computer Networks}}  \bibinfo{volume}{119}
  (\bibinfo{year}{2017}), \bibinfo{pages}{86--101}.
\newblock
\urldef\tempurl%
\url{https://doi.org/10.1016/j.comnet.2017.03.012}
\showURL{%
\tempurl}


\bibitem[Poisson et~al\mbox{.}(2024)]%
        {poisson2024gothx}
\bibfield{author}{\bibinfo{person}{Manuel Poisson}, \bibinfo{person}{Rodrigo
  Carnier}, {and} \bibinfo{person}{Kensuke Fukuda}.}
  \bibinfo{year}{2024}\natexlab{}.
\newblock \showarticletitle{GothX: a generator of customizable, legitimate and
  malicious IoT network traffic}. In \bibinfo{booktitle}{\emph{Proceedings of
  the 17th Cyber Security Experimentation and Test Workshop}}.
  \bibinfo{pages}{65--73}.
\newblock
\urldef\tempurl%
\url{https://doi.org/10.1145/3675741.36757}
\showURL{%
\tempurl}


\bibitem[S{\'a}ez-de C{\'a}mara et~al\mbox{.}(2023)]%
        {saez2023gotham}
\bibfield{author}{\bibinfo{person}{Xabier S{\'a}ez-de C{\'a}mara},
  \bibinfo{person}{Jose~Luis Flores}, \bibinfo{person}{Crist{\'o}bal Arellano},
  \bibinfo{person}{Aitor Urbieta}, {and} \bibinfo{person}{Urko Zurutuza}.}
  \bibinfo{year}{2023}\natexlab{}.
\newblock \showarticletitle{Gotham testbed: a reproducible IoT testbed for
  security experiments and dataset generation}.
\newblock \bibinfo{journal}{\emph{IEEE Transactions on Dependable and Secure
  Computing}} \bibinfo{volume}{21}, \bibinfo{number}{1} (\bibinfo{year}{2023}),
  \bibinfo{pages}{186--203}.
\newblock
\urldef\tempurl%
\url{https://doi.org/10.1109/tdsc.2023.3247166}
\showURL{%
\tempurl}


\bibitem[Saginbekov and Shakenov(2016)]%
        {saginbekov2016testing}
\bibfield{author}{\bibinfo{person}{Sain Saginbekov} {and}
  \bibinfo{person}{Chingiz Shakenov}.} \bibinfo{year}{2016}\natexlab{}.
\newblock \showarticletitle{Testing wireless sensor networks with hybrid
  simulators}.
\newblock \bibinfo{journal}{\emph{arXiv preprint arXiv:1602.01567}}
  (\bibinfo{year}{2016}).
\newblock
\urldef\tempurl%
\url{https://doi.org/10.48550/arXiv.1602.01567}
\showURL{%
\tempurl}


\bibitem[Sanchez et~al\mbox{.}(2014)]%
        {sanchez2014smartsantander}
\bibfield{author}{\bibinfo{person}{Luis Sanchez}, \bibinfo{person}{Luis
  Mu{\~n}oz}, \bibinfo{person}{Jose~Antonio Galache}, \bibinfo{person}{Pablo
  Sotres}, \bibinfo{person}{Juan~R Santana}, \bibinfo{person}{Veronica
  Gutierrez}, \bibinfo{person}{Rajiv Ramdhany}, \bibinfo{person}{Alex Gluhak},
  \bibinfo{person}{Srdjan Krco}, \bibinfo{person}{Evangelos Theodoridis},
  {et~al\mbox{.}}} \bibinfo{year}{2014}\natexlab{}.
\newblock \showarticletitle{SmartSantander: IoT experimentation over a smart
  city testbed}.
\newblock \bibinfo{journal}{\emph{Computer Networks}}  \bibinfo{volume}{61}
  (\bibinfo{year}{2014}), \bibinfo{pages}{217--238}.
\newblock
\urldef\tempurl%
\url{https://doi.org/10.1016/j.bjp.2013.12.020}
\showURL{%
\tempurl}


\bibitem[Shahid et~al\mbox{.}(2020)]%
        {shahid2020generative}
\bibfield{author}{\bibinfo{person}{Mustafizur~R Shahid},
  \bibinfo{person}{Gregory Blanc}, \bibinfo{person}{Houda Jmila},
  \bibinfo{person}{Zonghua Zhang}, {and} \bibinfo{person}{Herv{\'e} Debar}.}
  \bibinfo{year}{2020}\natexlab{}.
\newblock \showarticletitle{Generative deep learning for Internet of Things
  network traffic generation}. In \bibinfo{booktitle}{\emph{2020 IEEE 25th
  Pacific Rim International Symposium on Dependable Computing (PRDC)}}. IEEE,
  \bibinfo{pages}{70--79}.
\newblock
\urldef\tempurl%
\url{https://10.1109/PRDC50213.2020.00018}
\showURL{%
\tempurl}


\bibitem[Shelby et~al\mbox{.}(2022)]%
        {shelby2022rfc}
\bibfield{author}{\bibinfo{person}{Z. Shelby}, \bibinfo{person}{M. Koster},
  \bibinfo{person}{C. Bormann}, {and} \bibinfo{person}{P. van~der Stok}.}
  \bibinfo{year}{2022}\natexlab{}.
\newblock \bibinfo{title}{RFC 9176: Constrained RESTful Environments (CoRE)
  Resource Directory}.
\newblock
\urldef\tempurl%
\url{https://datatracker.ietf.org/doc/rfc9176/}
\showURL{%
Retrieved 2025-03-23 from \tempurl}
\newblock
\shownote{RFC Editor}.


\bibitem[Silmi et~al\mbox{.}(2020)]%
        {silmi2020wireless}
\bibfield{author}{\bibinfo{person}{Souhila Silmi}, \bibinfo{person}{Zouina
  Doukha}, \bibinfo{person}{Rebiha Kemcha}, {and} \bibinfo{person}{Samira
  Moussaoui}.} \bibinfo{year}{2020}\natexlab{}.
\newblock \showarticletitle{Wireless sensor networks simulators and testbeds}.
\newblock \bibinfo{journal}{\emph{arXiv preprint arXiv:2009.03640}}
  (\bibinfo{year}{2020}).
\newblock


\bibitem[Sowe et~al\mbox{.}(2014)]%
        {sowe2014managing}
\bibfield{author}{\bibinfo{person}{Sulayman~K Sowe}, \bibinfo{person}{Takashi
  Kimata}, \bibinfo{person}{Mianxiong Dong}, {and} \bibinfo{person}{Koji
  Zettsu}.} \bibinfo{year}{2014}\natexlab{}.
\newblock \showarticletitle{Managing heterogeneous sensor data on a big data
  platform: IoT services for data-intensive science}. In
  \bibinfo{booktitle}{\emph{2014 IEEE 38th international computer software and
  applications conference workshops}}. IEEE, \bibinfo{pages}{295--300}.
\newblock
\urldef\tempurl%
\url{https://doi.org/10.1109/compsacw.2014.52}
\showURL{%
\tempurl}


\bibitem[Tsakalidis et~al\mbox{.}(2023)]%
        {tsakalidis2023design}
\bibfield{author}{\bibinfo{person}{Sotirios Tsakalidis},
  \bibinfo{person}{George Tsoulos}, \bibinfo{person}{Dimitrios Kontaxis}, {and}
  \bibinfo{person}{Georgia Athanasiadou}.} \bibinfo{year}{2023}\natexlab{}.
\newblock \showarticletitle{Design and Implementation of a Versatile openHab
  IoT Testbed with a Variety of Wireless Interfaces and Sensors}. In
  \bibinfo{booktitle}{\emph{Telecom}}, Vol.~\bibinfo{volume}{4}. MDPI,
  \bibinfo{pages}{597--–610}.
\newblock
\urldef\tempurl%
\url{https://doi.org/10.3390/telecom4030026}
\showURL{%
\tempurl}


\end{thebibliography}

\appendix
\clearpage  % Forces the start of a new page in a two-column document

\section{STGen Usage Guideline}

This appendix contains the prerequisites \& installation process to run STGen.

\subsection{Prerequisites}
    \begin{itemize}
 
  \item 1: Python3 is also a prerequisite. You can check if Python is installed on your device by running the following command.
  \begin{lstlisting}[language=bash]
    $ python3
   \end{lstlisting}
\end{itemize}

\subsection{Installation to run in terminal mode}
    \begin{itemize}
  \item Step 1: Download the GitHub repository or clone it using the following command \footnote{https://github.com/rahmanmehraj182/Hybrid-IoT-Platform}.
  \begin{lstlisting}[language=bash]
    $ git clone git@github.com:rahmanmehraj182/Hybrid-IoT-Platform.git
   \end{lstlisting}
  \item Step 2: Open a terminal in your computer. Now, change the directory to run the server.
  \begin{lstlisting}[language=bash]
    $ cd STGen-Sensor-Traffic-Generator/launcher
   \end{lstlisting}

   \item Step 3: Run the Server.

  \begin{lstlisting}[language=bash]
    $ python3 STGen_server.py localhost 5004 5005 100
   \end{lstlisting}

  \item Step 4: Open another terminal tab and stay in the launcher directory. Now, run the Sensor(s).

  \begin{lstlisting}[language=bash]
    $ python3 STGen_launcher.py localhost 5004 200 temp:30:1 gps:10
   \end{lstlisting}

  \item Step 5: Open one more terminal tab and change the directory to run the client.
  \begin{lstlisting}[language=bash]
    $ cd STGen-Sensor-Traffic-Generator/iot/application
   \end{lstlisting}
   
   \item Step 6: Run the client.

  \begin{lstlisting}[language=bash]
    $ ./STGen_client -lclient1_sensor_log -slocalhost -rtemp_1 -p5005
   \end{lstlisting}

\end{itemize}

\subsection{Installation to run from WebUI}
   System Requirements: Node.js 18.17 or later.

\begin{itemize}
  \item Step 1: Download the GitHub repository or clone it using the following command \footnote{https://github.com/subratanath123/STGen-A-Novel-Lightweight-Sensor-Traffic-Generator-Web-UI}.
  \begin{lstlisting}[language=bash]
    $ git clone https://github.com/subratanath123/STGen-A-Novel-Lightweight-Sensor-Traffic-Generator-Web-UI
   \end{lstlisting}
  \item 2: Open a terminal on your computer. Now, change the directory to the downloaded codebase.
  \begin{lstlisting}[language=bash]
    $ cd STGen-A-Novel-Lightweight-Sensor-Traffic-Generator-Web-UI
   \end{lstlisting}

   \item Step 3: Run the WebUI.

  \begin{lstlisting}[language=bash]
    $ npm run dev
   \end{lstlisting}

    \item Step 4: Open another terminal window. 

    \item Step 5: Download the WebUI backend GitHub repo or clone it using the following command \footnote{https://github.com/subratanath123/STGen-A-Novel-Lightweight-Sensor-Traffic-Generator-Backend-Launcher}.
      \begin{lstlisting}[language=bash]
    $ git clone https://github.com/subratanath123/STGen-A-Novel-Lightweight-Sensor-Traffic-Generator-Backend-Launcher
   \end{lstlisting}
   \item Step 6: Change the directory to the downloaded codebase.  
  \begin{lstlisting}[language=bash]
    $ cd STGen-A-Novel-Lightweight-Sensor-Traffic-Generator-Web-UI
   \end{lstlisting}

   \item Step 7: Run the WebUI Backend
  \begin{lstlisting}[language=bash]
    $ ./gradlew bootRun
   \end{lstlisting}

\end{itemize}

\end{document}